# Characterizing Fill Factor Limitations in Perovskite-Silicon Tandem Solar Cells

Yueming Wang[*,1], Nan Sun[1], Chris Dreessen[1], Gaosheng Huang[1], Alexander Eberst[1], Kaining Ding[1], and Thomas Kirchartz[*,1,2]

[1]IMD-3 Photovoltaics, Forschungszentrum Jülich, 52425 Jülich, Germany
[2]Faculty of Engineering and CENIDE, University of Duisburg-Essen, Carl-Benz-Str. 199, 47057 Duisburg, Germany
E-mail: yue.wang@fz-juelich.de
E-mail: t.kirchartz@fz-juelich.de

## Abstract

Perovskite-silicon tandem technology has exceeded the single junction theoretical efficiency limit. However, there is still distance to the thermodynamic limit mainly caused by the fill factor. This work presents a methodology to illustrate the mechanisms of *FF* loss in perovskite-Si monolithic tandem solar cells. Apart from the series resistance related loss characterized by electroluminescence, another loss factor is from the photoshunt, a phenomenon in which the parallel resistance apparently reduces under illumination in perovskite solar cells due to the moderate charge transport layer mobility. In addoition, the two-diode property of the Si cell can also influence the *FF* of tandem devices. The photoshunt can be hidden when the bottom cell is over illuminated, which explains highly efficient tandem solar cells are usually bottom cell limited. This work outlines strategies that overcoming the photoshunt issue can move the perovskite top cell closer to low *FF* losses in tandem solar cells.

## 1. Introduction

Perovskite-silicon tandem solar cells have recently overcome the single-junction detailed balance efficiency limit of 33%[1,2] and are rapidly increasing the efficiency difference to the respective perovskite and silicon single-junction solar cells.[3,4,5,6] Thus, given future improvements in device stability, the long quest[7,8] for achieving reasonably cost-efficient and high-efficiency multijunction solar cells might eventually become a reality. In the pursuit of even higher efficiencies, significant research has been devoted to improving the open-circuit voltages of the ~1.7 eV bandgap perovskite subcells.[3,9] However, if we study the efficiency losses in tandem cells relative to their thermodynamic limit (see Figure 1), we note that the most significant losses of current tandem solar cells are in the fill factor. Furthermore, the most significant improvement in the current record tandem solar cell relative to many other recently published tandem solar cells is its improved fill factor.[5,6] Unfortunately, the fill factor is the least intuitive of the three classical photovoltaic performance parameters, short-circuit current density $J_{sc}$, open-circuit voltage $V_{oc}$, and fill factor $FF$. This is because the $FF$ describes the relative performance of the cell at the maximum power point, where all efficiency-limiting photovoltaic processes are important. Thus, the $FF$ is affected by resistive effects (series and parallel resistance), recombination dynamics (with an ideality factor $n_{id}$), and inefficient charge-carrier collection. In the specific case of multijunction solar cells, an additional difficulty is that the exact current matching condition between the subcells of the multijunction will have a strong effect on the $FF$. Thus, the highest $FF$s in multijunction solar cells can be achieved by the poorest current matching. This unfortunately defies an important feature of the $FF$ as a quality metric in single-junction solar cells, namely, that higher $FF$s are always better. A final complication, specific to halide perovskites, is their ionic-electronic conductivity, which leads to the transient nature of all photovoltaic performance parameters, primarily the $FF$.

The maximum power point of any photovoltaic device is eventually the only bias condition that matters for quantifying performance. Thus, it is of paramount importance to have a sound understanding of the fill factor and efficiency-limiting mechanisms that have a particularly strong influence on the fill factor for the development of tandem solar cells. The present study starts with the thermodynamic limits of efficiency and fill factor in tandem solar cells and then focuses on the specific losses that are present in the perovskite and silicon subcells. Perovskite subcells can suffer from a phenomenon that we refer to as photoshunt. This means that the illuminated current-voltage (*J-V*) curves are affected in a way similar to a reduced shunt resistance, but this phenomenon is not visible in the dark. The photoshunt is caused by inefficient charge carrier extraction at short circuit and low forward voltages, which leads to additional recombination under illumination. In perovskite solar cells, the slow charge carrier extraction speed is usually directly dependent on the limited mobility of the charge transport materials. The solar cells were simulated using an equivalent circuit model to understand the effects of the mobility-dependent photoshunt on the *FF*s and efficiencies of single-junction perovskite solar cells, which were then extended to perovskite-Si monolithic tandem devices. We also studied how the ideality factors of the Si bottom cells affect the *FF*s of the tandem devices. A common method used to analyze the *FF* loss of perovskite-Si monolithic tandem solar cells is the extraction of pseudo-*JV* curves by electroluminescence measurements. However, the EL characterization method can only be used to analyze the *FF* losses from the series resistance. This work expands the understanding of the mechanism of *FF* losses in monolithic tandem solar cells from series resistance to photoshunt, which is highly relevant to the properties of charge transport materials.

## 2. Efficiency Losses in Tandem Solar Cells

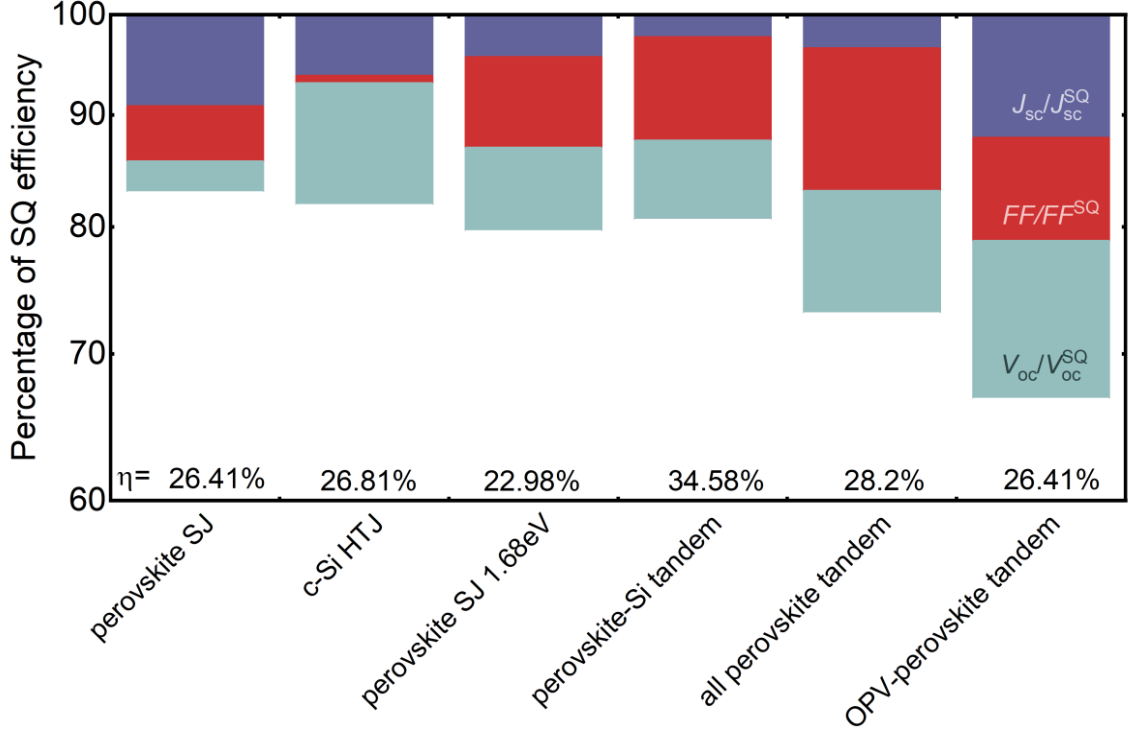

**Figure 1.** The relative losses of photovoltaic parameters of a selected group of high-efficiency single-junction and tandem solar cells are presented, normalized to their respective values within the framework of the SQ model. As the product of each ratio gives the normalized efficiency according to equation (1), and because multiplicative factors preserve their area on a logarithmic scale, the y-axis is a logarithmic axis in % of losses, whereby the lowest edge of each column represents the normalized efficiency $\eta^{\text{real}}/\eta^{\text{SQ}}$.

To quantify the losses in the classical photovoltaic parameters, it is useful to compare them to their maximum value, which we can assume to be the value in the so-called Shockley-Queisser model. The PV solar-to-electrical power conversion efficiency can be mathematically expressed as $\eta = (J_{\text{sc}} V_{\text{oc}} FF)/P_{\text{in}}$. Dividing each quantity by their respective maximum value in the SQ model, yields a normalized version of this equation[10, 11]

$$\frac{\eta^{\text{real}}}{\eta^{\text{SQ}}} = \frac{J_{\text{sc}}^{\text{real}}}{J_{\text{sc}}^{\text{SQ}}} \frac{V_{\text{oc}}^{\text{real}}}{V_{\text{oc}}^{\text{SQ}}} \frac{FF^{\text{real}}}{FF^{\text{SQ}}}. \tag{1}$$

Equation (1) can be used to compare the efficiency losses in the recorded single-junction and tandem solar cells. These losses can be divided into three factors according to Equation (1) and as shown in Figure 1. The blue sections of the bars in Figure 1 represent the loss due to imperfect photocurrent

$J_{sc}/J_{sc}^{SQ}$. The photocurrent of a solar cell is highly dependent on the light absorptance of the photovoltaic materials, light management, and the mobility of the charge transport layers. In addition to the factors mentioned above, current matching is important for reducing current loss in monolithic tandem solar cells. The single-junction perovskite solar cell with record efficiency still has 9.2% efficiency loss from photocurrent.[12] The highest efficient single-junction perovskite solar cell with 1.68 eV bandgap, which is frequently used for tandem applications with Si cells, has efficiency loss from photocurrent less than 5%.[13] For the record crystalline Si heterojunction solar cell, the efficiency loss from photocurrent is 6.23%.[14] The record two-terminal perovskite-Si tandem cell[6] and all perovskite tandem cell[15] both have very low current losses of 2.35% and 3.49%, respectively. The organic (OPV)-perovskite tandem cell[16] has a very high current loss of 12.16% compared to other types of tandem- and single-junction solar cells.

The losses in the open-circuit voltage $V_{OC}$ can be divided into two parts: (1) the difference between the actual shape quantum efficiency and the ideal step-function assumed in the Shockley-Queisser (SQ) model, which is given by $V_{oc}^{rad}/V_{oc}^{SQ}$, and (2) the voltage loss from nonradiative recombination, which is given by $V_{oc}^{real}/V_{oc}^{rad}$.[11] The dark green areas in Figure 1 show the total open-circuit voltage ($V_{OC}$) losses in different types of solar cells. The record-efficiency single-junction perovskite solar cell[12] has the lowest efficiency loss from $V_{OC}$ of 2.74% among the different types of photovoltaic devices, as shown in Figure 1. The $V_{OC}$ loss of the tandem cells is the sum of the $V_{OC}$ losses of each subcell. Both all-perovskite and OPV-perovskite tandem cells have about 10% efficiency losses from $V_{OC}$, which is a bit higher than the 7% of perovskite-Si tandem solar cell.[6, 15, 16]

The fill factor (*FF*) loss of a solar cell can also be divided into two parts.[17] Under ideal conditions, a solar cell is considered a perfect diode with an ideality factor of 1 and no resistive losses; the maximum possible *FF* is then only a function of $V_{OC}$. The *FF* loss due to $V_{OC}$ loss is written as $FF_0(V_{oc}^{real})/FF_0(V_{oc}^{SQ})$. The second set of phenomena that cause *FF* losses in solar cells are the resistive effects. The resistances can be in series or in parallel with the diode and current source. The resistive *FF* loss is described as $FF^{real}/FF_0(V_{oc}^{real})$. The dark red areas in Figure 1 represent the total *FF* losses in the single-junction and tandem solar cells.

The reported single-junction perovskite solar cell with record efficiency has a lower *FF* loss (7.86%) than the highest efficient 1.68 eV single-junction perovskite solar cell (11.83%). The record crystalline Si heterojunction solar cell has a very low *FF* loss because of the Auger limit.[18] The three different types of record tandem cells in figure 1 all have significant efficiency losses from *FF*, 10.72% for perovskite-Si tandem cell,[6] 13.44% for all perovskite tandem cell[15] and 9.01% for OPV-perovskite tandem cell.[16]

The *FF* of tandem cells is influenced not only by $V_{OC}$s and resistive effects but also by the current matching of each subcell. In monolithic tandem solar cells, the *FF*s increase with the current mismatch of the sub-cells.[19] For highly efficient 2-terminal perovskite-Si tandem cells, the Si bottom subcells usually limit the photocurrent under AM1.5G illumination. This bottom cell limitation is usually beneficial also for achieving higher *FF*s.[5, 6]

### 2.1. Thermodynamic Limit of Fill Factor in Tandem Solar Cells

The Shockley-Queisser (SQ) model is one of the most important theories for describing the limits of an idealized single-junction solar cell[2] and its general framework can be extended to situations such as multijunctions,[20, 21] up- and down-conversion,[22, 23] or impact ionization[24] in a relatively straightforward way. This feature makes it extremely useful to study concepts (termed "third generation photovoltaics" by Martin Green)[25] that go beyond the classical single junction solar cell that was the topic of the original 1961 paper. The SQ model is based on the radiation balance between the sun, the solar cell, and its environment (Earth).[1] The three bodies are quantified only by their temperature (the temperature of the solar cell and Earth is assumed to be identical) and, in the case of the solar cell, by its absorption onset which thereby becomes the main variable of the model. The assumptions used to create a mathematical framework for calculating the efficiency within the SQ model are as follows:

When light enters the absorber layer of a solar cell, photons with energy $E$ above the bandgap energy $E_g$ of the absorber layer can be absorbed, whereas photons with energy lower than $E_g$ do not interact with the photovoltaic absorber at all. Thus, the absorptance of the solar cell is a Heaviside function with the bandgap as the position of the step on the energy axis. Every absorbed photon with energy $E > E_g$ generates an electron-hole pair. i.e. processes such as free-carrier absorption or multiple-exciton generation are excluded. Every generated electron-hole pair thermalizes to the same average energy in thermal equilibrium with the solar cell. All photogenerated and thermalized charge carriers must be collected at the electrode of the solar cell. There is no other kind of recombination happening in the solar cell apart from radiative recombination.[2]

The short-circuit current density of a solar cell in the SQ limit is given by

$$J_{sc,\,SQ} = q \int_{E_g}^{\infty} \phi_{sun}(E) dE \qquad (2)$$

The dark saturation current or, in another word, the solar cell's thermal emission at room temperature (300K) is (expressed as a current density)

$$J_{0,SQ} = q \int_{E_g}^{\infty} \phi_{bb}(E) dE \qquad (3)$$

Here, $\phi_{bb}$ is the photon flux of a black body at room temperature emitted into the $2\pi$ halfsphere above the cell which is described as $\phi_{bb}(E) = \frac{2\pi E^2}{h^3 c^2} \frac{1}{\left[\exp\left(\frac{E}{kT}\right)-1\right]} \approx \frac{2\pi E^2}{h^3 c^2} \exp\left(\frac{-E}{kT}\right)$. $h$ is Planck's constant, $E$ is the photon energy, and $c$ is the speed of light in vacuum. $\phi_{sun}$ is the solar spectrum. Whereas the original SQ study used a black body spectrum at 6000K, we will use the standardized AM1.5G spectrum for better comparability with standard testing conditions.

The current voltage (*J-V*) curve of a solar cell at SQ limit is given by equation (4) with the ideality factor $n_{id} = 1$, and $kT$ is thermal energy at 300K.

$$J = J_{0,SQ}\left(\exp\left(\frac{qV}{n_{id}kT}\right) - 1\right) - J_{sc,SQ} \qquad (4)$$

The open-circuit voltage of a solar cell within the framework of the SQ model is obtained by solving Equation (4) and is expressed via

$$V_{oc,\,SQ} = \frac{n_{id}kT}{q} \ln\left(\frac{J_{sc,\,SQ}}{J_{0,SQ}} + 1\right) \qquad (5)$$

The *FF* of a solar cell within the framework of the SQ model is calculated as equation (6) with $v_{oc} = qV_{oc,\,SQ}/kT$.

$$FF_{SQ} = \frac{v_{oc} - \ln(v_{oc} + 0.72)}{v_{oc} + 1} \qquad (6)$$

The maximum efficiency of a single junction solar cell in the SQ model for an AM1.5 G spectrum is about 33% at an optimum bandgap.[1] To increase efficiencies beyond the SQ limit for a single junction solar cell, research has focused on fabricating multijunction devices that provide the arguably most straightforward path to overcome the limitations of the original SQ model.[26, 27] The principle of tandem photovoltaic operation is arranging solar cells with different bandgaps optically in series to reduce thermal energy loss of photogenerated carriers and achieve efficient spectral division.[28] Usually, the higher bandgap material is on top of lower bandgap material to maximize solar spectrum utilization.[29]

The two-junction solar cell with two-terminal (2-T) electrical contacts is the most actively studied tandem device structure now. Here the two subcells are connected in series both optically and electrically.[3, 5, 6, 30, 31, 32] In series connected 2-T device configuration, one electron and one hole need to be collected at their respective junctions while one recombination process needs to happen at the recombination junction that connects the 2 sub-cells.[33, 34] Each sub-cell takes its own part of the solar spectrum as input.[33] Therefore, the theoretical efficiency limit of monolithic 2-T tandem solar cells can be studied in the framework of the SQ model by assuming that the sub-cells are ideal and match the assumptions of the single junction SQ model with the one exception that the bottom cell only absorbs the part of the spectrum between the two band gaps.[35]

Figure 2b shows the maximum values of $J_{sc}$ in the monolithic 2-T tandem solar cells as a function of the two bandgaps of the two subcells. The ideal $J_{sc}$ reached by a perfectly current-matched tandem solar cell is half of the ideal $J_{sc}$ of the lower-bandgap sub cell. Otherwise, the photocurrent of tandem cells is limited by the subcell producing the lower short circuit current density. The $V_{oc}$s of tandem cells are equal to the sum of each sub-cells' open-circuit voltages as shown in Figure 2c. At least conceptually and to a good approximation also practically, both the $J_{sc}$ and $V_{oc}$ of tandem cells result in a straightforward manner from the values of the individual subcells. However, the *FF* of a tandem solar cell is not a simple function of the *FF*s of its subcells. Even without considering the influence of resistance and the non-ideal ideality factor, current matching plays an important role in the *FF* of a tandem device. As shown in Figure 2d, the lowest *FF*s occurred for the condition of perfect current matching of each subcell. The larger the current mismatch of the two subcells, the higher the *FF*. The potential efficiencies of monolithic 2-T tandem solar cells are predicted in Figure 2a, which is strongly dependent on the bandgap of the subcells.

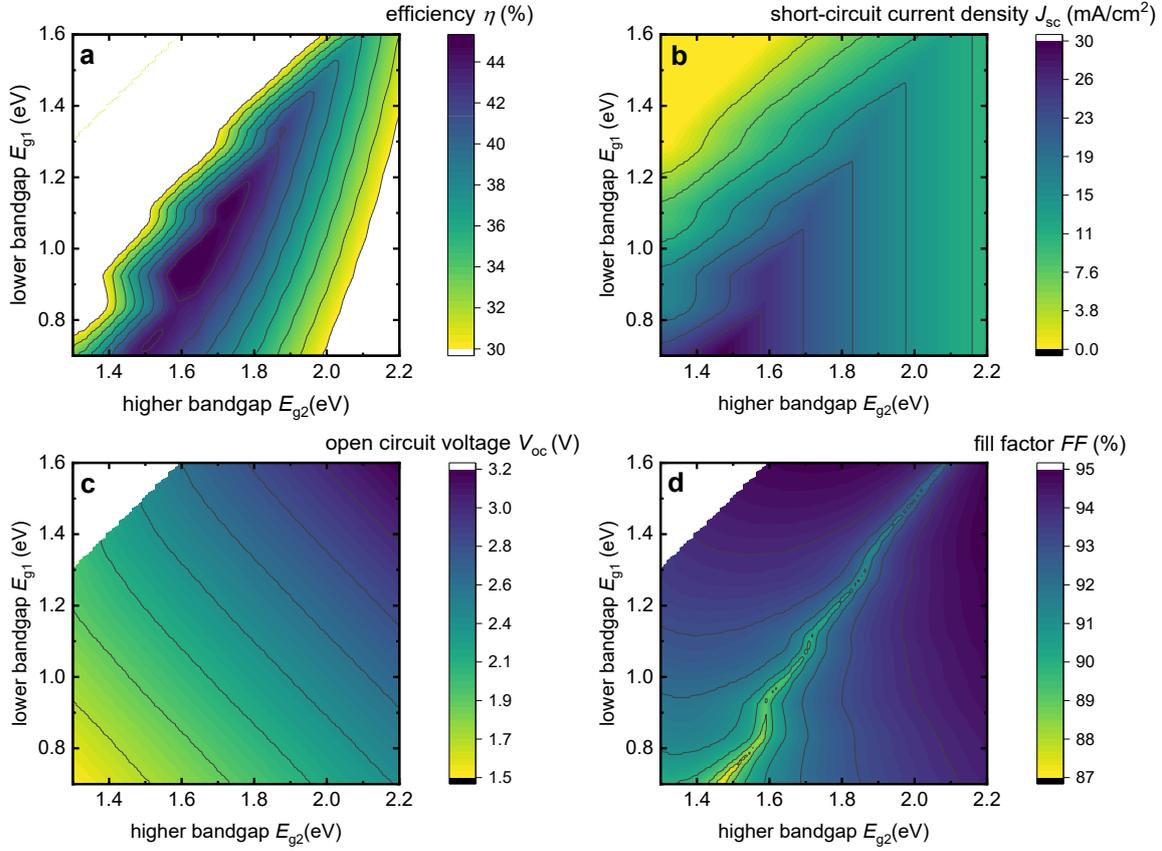

**Figure 2.** Contour plots for the photovoltaic parameters in the radiative limit for (a) efficiencies, (b) short-circuit current densities, (c) open-circuit voltages, and (d) fill factors of monolithic 2-T tandem solar cells. All data are plotted as a function of the two band gaps of the two subcells.

### 3. Electroluminescence and Pseudo-*JV* Curves of Single-Junction and Tandem Solar Cells

Currently, the most widely applied solar photovoltaic technology is based on crystalline silicon (c-Si).[36] Potential partner sub-cells for c-Si in tandem photovoltaic are hybrid organic inorganic metal halide perovskite solar cells.[37, 38] The most suitable bandgap of perovskite is approximately 1.7 eV, combining with the fixed bandgap of c-Si 1.12 eV, as shown in Figure 2. Figure 2 shows the theoretical efficiency limit of monolithic perovskite-Si tandem solar cells for the AM1.5G spectrum at 300 K. However, in real devices, resistive effects cannot be avoided and usually are important reasons for efficiency losses.[17, 39, 40] It is already clear that the series resistance $R_s$ has a strong impact on the *FF*s of single junction solar cells.[17, 41] However, in tandem devices, the parallel resistance $R_p$ has unignorable influence on $FF$s.

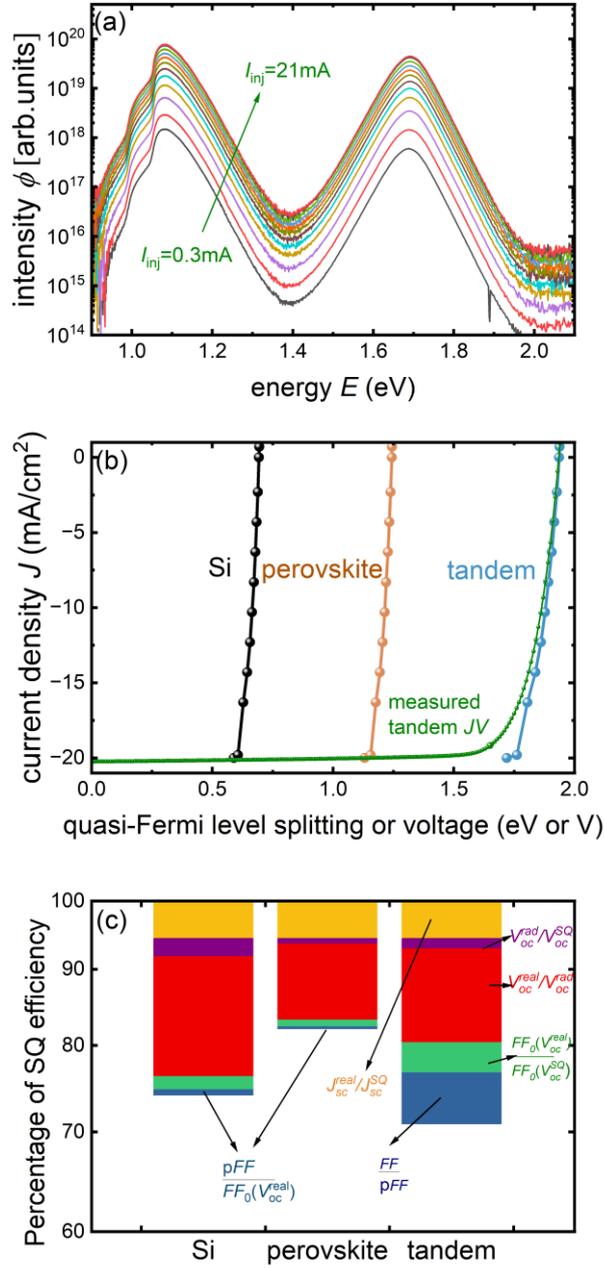

**Figure 3.** (a) Injection-dependent absolute electroluminescence (EL) spectra of the perovskite and silicon sub-cells. Measurements were performed under various injected current to vary bias voltages. (b) reconstructed *J-V* curves calculated from injection-dependent electroluminescence measurements (sphere) The *J-V* measurement under 1-sun illumination of tandem cell is plotted in green. (c) the percentage of Shockley-Queisser efficiencies for perovskite-Si tandem solar cell and sub-cells in losses of current densities, voltages and fill factors corresponding to a multiplicative factor in equation (8) plotted on a logarithmic scale.

Electroluminescence (EL) is a commonly used method [42, 43, 44, 45] on multijunction solar cell characterization as it allows to access the Fermi-level splittings of each individual subcell as a function of the injection current density without the need to actually create extra contacts between the different subcells. In this context, EL offers access to the internal voltages of the subcells purely optically via the relationship between luminescence emission and Fermi-level splitting.[46, 47] The EL measurements offer the possibility of gaining deep insights into the sub-cell limitations of an actual tandem cell. In Figure 3, we identify the efficiency loss from the series resistance in a perovskite-Si solar cell with an efficiency of approximately 31%. The perovskite-Si tandem solar cells were based on triple halide perovskite $Cs_{0.22}FA_{0.78}PbI_{0.85}Br_{0.15}$ with 5% $MAPbCl_3$ and 0.7% $Cl-FAPbCl_3$ top cells and silicon heterojunction bottom cells with random pyramids on both the front and rear sides. The fabrication details of the tandem solar cells are provided in the Supporting Information. The photovoltaic parameters of the 18 tandem devices are summarized in Figure S3. The *J-V* curves of a champion device are plotted in Figure S4. The EL spectra of the Si and perovskite sub-cells at different injected currents (ranging from 0.3 mA to 21 mA) are shown in Figure 3a. The injected currents flow equally through the two sub-cells because of the monolithic structure of the tandem device. The internal voltage of the Si and perovskite sub-cells at different injected currents can be quantified by a combination of the external quantum efficiency and EL spectrum via Equation (7). The details of the data analysis are provided in the Supporting Information.

$$\phi_{\mathrm{em}}(E) = Q_{\mathrm{e}}(E)\phi_{\mathrm{bb}}(E)\left[\exp\left(\frac{qV}{kT}\right) - 1\right] \quad (7)$$

Here, $\phi_{\mathrm{em}}(E)$ is the EL emission as a function of photon energy, $Q_{\mathrm{e}}(E)$ is external quantum efficiency of the solar cell, $\phi_{\mathrm{bb}}(E)$ is the black body photon flux in units of photons per area and time, *V* is the internal voltage on each sub-cells, and $kT/q$ is the thermal voltage (25.9 mV at room temperature). The pseudo *J-V* curves of the tandem cell and sub-cells were reconstructed by calculating the internal voltages for each subcell at each injected current and adding the internal voltages up to arrive at the pseudo-*JV* curve of the tandem solar cell. The results are shown in Figure 3b. The efficiency losses of the tandem cell and sub-cells (Figure 3c) were calculated from the SQ model, as described in Figure 1 and Section 2, but more specifically as follows

$$\frac{\eta_{\mathrm{real}}}{\eta_{\mathrm{SQ}}} = \frac{J_{\mathrm{sc}}}{J_{\mathrm{sc}}^{\mathrm{SQ}}} \frac{V_{\mathrm{oc}}^{\mathrm{rad}}}{V_{\mathrm{oc}}^{\mathrm{SQ}}} \frac{V_{\mathrm{oc}}^{\mathrm{real}}}{V_{\mathrm{oc}}^{\mathrm{rad}}} \frac{FF_0(V_{\mathrm{oc}}^{\mathrm{real}})}{FF_0(V_{\mathrm{oc}}^{\mathrm{SQ}})} \frac{FF_{\mathrm{real}}}{FF_0(V_{\mathrm{oc}}^{\mathrm{real}})} \quad . \quad (8)$$

The calculation details of the radiative voltage are provided in the Supporting Information. The *FF* loss analysis method is the same as that in refs.[17, 48] The *FF* losses from the series resistance can be analyzed from the EL characterization (the difference between *FF* and pseudo-*FF*). However, studying only electroluminescence is insufficient to understand the *FF* losses of tandem solar cells, as shown in Figure 3c.

## 4. Effect of the Photoshunt on Single-Junction and Tandem Solar Cell Fill Factors

Halide perovskite solar cells typically have very high shunt resistances, $R_p$, in the dark that do not have a strong effect on the fill factor of these devices which is described in the *J-V* curves in Figure S7b. However, as is the case for most thin-film solar cells,[41, 49, 50] the current-voltage curves under illumination show a linear regime at low forward bias that appears similar to a shunt resistance, but with an entirely different magnitude than in the dark (shown in Figure S7c). These apparent photoshunts, $R_{p,photo}$, are recombination currents that appear linear with the external voltage and therefore resemble a shunt. The linearity with external voltage seems to be entirely at odds with the fact that recombination currents are generally exponentially dependent on quasi-Fermi level splitting. However, as shown in ref[41]., the Fermi-level splitting and the external voltage at low to moderate forward bias are so completely decoupled that recombination can be both linear in external voltage and exponential in Fermi-level splitting at the same time. The requirement for this to happen are strong gradients of the quasi-Fermi level at short circuit and moderate forward bias that will occur primarily in those layers that have low conductivities, that is in perovskite solar cells primarily the transport layers. As these apparent photoshunts will also occur in perovskite subcells used for tandem cells, it is important to investigate their effect on the overall tandem fill factor and efficiency.

While photoshunts have a somewhat complicated to explain origin, their mathematical description is conveniently simple. For a given illuminated current-voltage curve, photoshunts are phenomenologically indistinguishable from a dark shunt, which implies that we can use the same framework created to study the effect of a dark shunt on the *FF*. Solar cells can be modelled by an equivalent circuit including one or more diodes, as well as elements representing the generation and losses in devices when they are illuminated. Here we simulate the monolithic perovskite-Si tandem solar cell with an equivalent circuit model given by

$$J = J_0 \left(\exp\left(\frac{q(V-JR_s)}{n_{id}kT}\right) - 1\right) + \frac{(V-JR_s)}{R_{p,photo}} - J_{sc} \qquad (9)$$

to understand the effect of $R_{p,photo}$ on devices performance especially *FF*s and efficiencies. To make the simulation results closer to real devices, we use the *J-V* curve and external quantum efficiency of the heterojunction Si solar cell with record efficiency[14] but not a solar cell with a bandgap of 1.12 eV in ideal SQ model. The perovskite top cells are approximated by the modified diode equation with varying bandgaps and parallel resistances. To isolate the effect of $R_{p,photo}$ on solar cell performance, here the $R_s$ is assumed as 0.

Figure 4a shows the efficiency of monolithic perovskite-Si tandem solar cells changing with parallel resistance and bandgaps of perovskite bottom cell intuitively. Apart from the situation that $R_{p,photo}$ is very low (less than 100 Ωcm²), the efficiency of tandem devices shows a maximum point with the bandgap of top cells changing. The maximum points keep moving to higher bandgap with the $R_{p,photo}$ increasing. The matching situation plays a significant role in the efficiencies and *FF*s of tandem cells. Similar as the results in Figure 2d, *FF* of tandem devices (shown in Figure 4b) reach minimal points when top cells perfectly match (about 1.74 eV) with the fixed bandgap bottom cell but only in the case that $R_{p,photo}$ is not very low (higher than $10^3$ Ωcm²). When the $R_{p,photo}$ is at very low value (lower than 400Ωcm²), the *FF*s of tandem cells continue decrease with bottom cell $E_g$ going higher but decrease slower after the perfect matching point. The best efficiency of tandem cells is not reached under perfectly matched conditions but always under a slight bottom cell limitation.

Figure 4c and d show the efficiencies and *FF*s vs. bandgaps for the single-junction perovskite solar cells that were applied as top cells in tandem devices in Figure 4a and b when varying the parallel resistances. The efficiencies of single junction solar cells are basically followed by equation (9) which are strongly dependent on $R_{p,photo}$ and weakly influenced by $E_g$. When $R_{p,photo}$ is sufficiently high, the *FF*s go saturation basically constant near the ideal values.

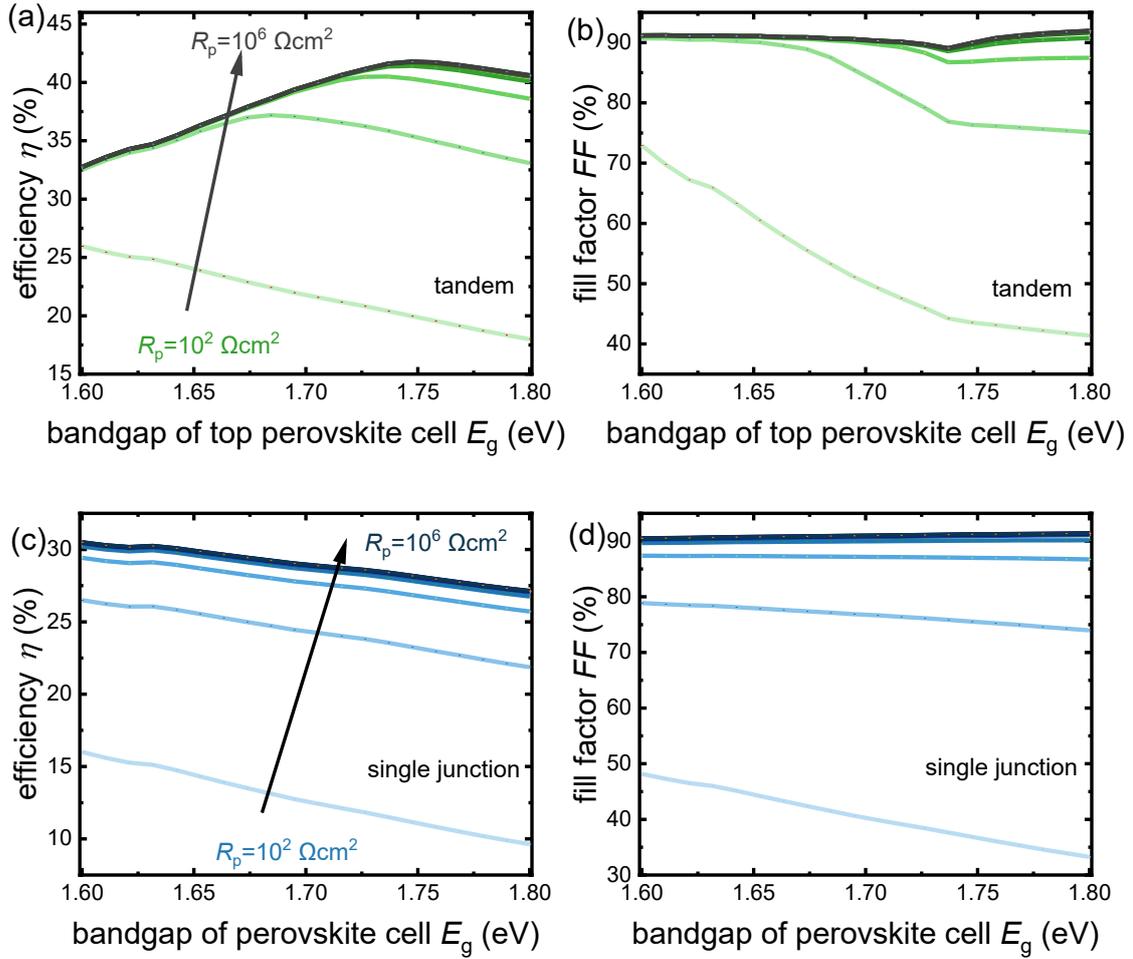

**Figure 4.** (a) Efficiency and (b) *FF* vs. bandgap for tandem solar cells with record heterojunction c-Si fixed as the bottom cell and perovskite top cells varying parallel resistance. (c) Efficiency and (d) *FF* vs. bandgap for different values of the parallel resistance of the single-junction perovskite solar cells.

To further understand how parallel resistance influences the performance of tandem devices, the shifted illuminated *J-V* curves (described in equation (10)) in the first quadrant of tandem devices and single-junction perovskite solar cells with different $R_{p,photo}$ values were plotted in Figure 5a and b based on the simulation results. The simulation model is the same as that used in Figure 4, in which the perovskite top cells are approximated by the modified one-diode model with a bandgap of 1.74 eV connected in series with the heterojunction Si solar cell with record efficiency. When $R_{p,photo}$ decreased from $10^6$ Ωcm² to $10^2$ Ωcm², the photoshunt in the *J-V* curves of both the tandem devices (Figure 5a) and single-junction perovskite solar cells (Figure 5b) became more obvious.

$$J_{shift} = J(V) + J_{sc} \qquad (10)$$

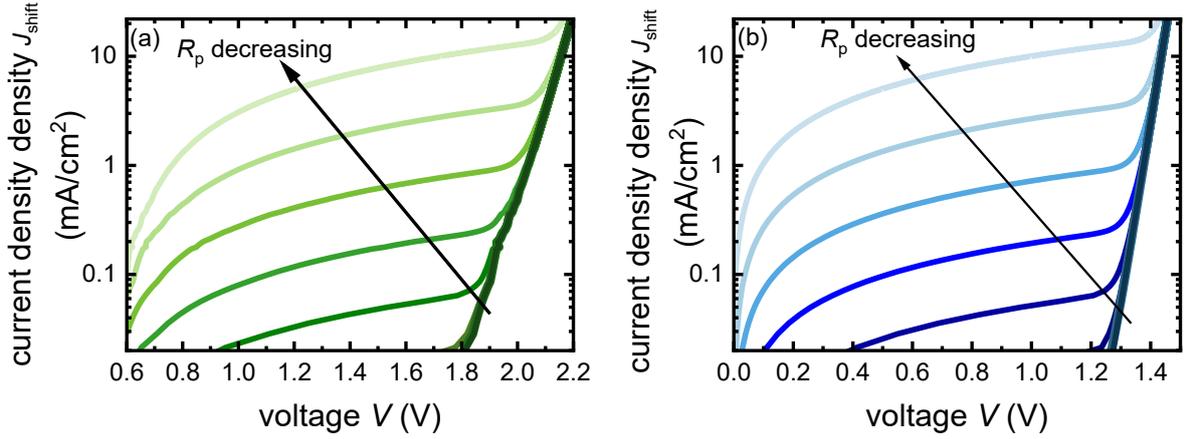

**Figure 5**. Simulated *J-V* curve under 1 sun illumination of (a) perovskite-Si tandem solar cells and (b) single-junction perovskite solar cells. *J-V* curve shifted to the first quadrant by adding $J_{sc}$ with varied parallel resistances $R_{p,photo}$ (from $10^2$ to $10^6$ $\Omega cm^2$). The bandgap of perovskite is 1.74 eV.

We previously discussed that the most likely origin of the apparent photoshunts are the low conductivities of the generally undoped electron and hole transport layers that are used in most perovskite (p-i-n-based) single junction and tandem solar cells.[17, 49] Thus, a logical question is how to characterize and quantify these gradients in Fermi-level splitting and its origin, namely the slow transport through the transport layers.

We therefore introduced a convenient metric to describe the speed of transport through ETL and HTL.[17, 51, 52] The resulting parameter, the exchange velocity $S_{exc}$ describes how fast transport of electrons through the electron transport layer (ETL) and holes through the hole transport layer (HTL) is, and at steady state, the $S_{exc}$ can be extracted from the voltage-dependent photoluminescence (PL) measurements via[53]

$$S_{\text{exc}} = J/\left\{qn_0\left[\exp\left(\frac{qV_{\text{ext}}}{2kT}\right) - \exp\left(\frac{qV_{\text{int}}}{2kT}\right)\right]\right\} \tag{11}$$

Equation (11) describes a phenomenological definition of $S_{exc}$ that is based on the three observables external voltage $V_{ext}$, internal voltage $V_{int}$ (via the PL), and externally measured current density $J$. However, we can also derive an equation for $S_{exc}$ based on simplified assumptions regarding transport in the transport layers (see e.g. equation (12) in ref. [54]). In the limit of having an electrostatic voltage drop $\Delta\varphi$ over the transport layers that is much larger than thermal voltage $kT/q$, we find that $S_{exc} = \mu_{TL}F$, where $F$ is the electric field inside the transport layers (assumed equal in ETL and HTL), $\mu_{TL}$ is the mobility of charge transport layer. Thus, the exchange velocity is simply the drift velocity of electrons in the ETL or holes in the HTL. Both velocities must be assumed to be approximately equal for this simple analytical treatment to work. If we stick to this simple symmetrical treatment of ETL and HTL and assume that the applied voltage drops equally over ETL and HTL and, due to ionic screening, not over the absorber layer, then we obtain $F = (V_{bi,TL} - V/2)/d_{TL}$, where $V_{bi,TL}$ is the electrostatic potential difference that drops over one transport layer at short circuit. If we combine this with our recently derived equation[51] for a current-voltage curve of a solar cell, where charge extraction is limited by the finite mobility of electrons and holes in the transport layers, we arrive at

$$J = \left(\frac{1}{1 + \frac{d_{abs} d_{TL}}{\tau_{abs} \mu_{TL}(V_{bi,TL} - V/2)}}\right) \left(\frac{q d n_0}{\tau_{abs}} \left[\exp\left(\frac{q V_{ext}}{2 k_B T}\right) - 1\right] - J_{gen}\right) \quad (12)$$

which connects the material and device properties of transport layers (mobility $\mu_{TL}$, thickness $d_{TL}$) and the absorber layer (lifetime $\tau_{abs}$, thickness $d_{abs}$) with the external voltage $V_{ext}$, the maximum photocurrent $J_{gen}$ and the measured current density $J$. The first term in brackets in equation (12) corresponds to a collection efficiency that describes the likelihood of extracting a charge carrier generated by light absorption in the absorber layer as a function of how quickly the charges are transported through ETL and HTL (depending on $\mu_{TL}$) and how quickly they recombine in the absorber layer ($\tau_{abs}$). The result further depends on the electric field (more field leads to faster extraction) as well as on the relative thicknesses of absorber and transport layers. Higher thicknesses of both the absorber and transport layer make charge extraction more difficult.

By taking the derivative $dV/dJ$ of the current-voltage curve as given by equation (12) at short circuit, we can now derive an analytical equation for the photoshunt in the situation, where the finite mobility $\mu_{TL}$ of the transport layers limits charge extraction. The resulting equation for the photoshunt resistance $R_{p,photo}$ in units of $\Omega$cm² is given by

$$R_{\text{p}} = \left.\frac{dV}{dJ}\right|_{V=0} = \frac{2\left(1 + \frac{d_{\text{abs}}d_{\text{TL}}}{\mu_{\text{TL}}\tau_{\text{abs}}V_{\text{bi,TL}}}\right)^2 \mu_{\text{TL}}\tau_{\text{abs}}V_{\text{bi,TL}}^2}{d_{\text{abs}}d_{\text{TL}}J_{\text{gen}}} \qquad (13)$$

and the result is illustrated in Figure 6 as a function of the transport layer mobility $\mu_{\text{TL}}$ and using the absorber layer lifetime $\tau_{\text{abs}}$ as a parameter. We note that the photoshunt is not a monotonous function of the transport layer mobility but instead features a minimum at a finite mobility. This observation is consistent with the situation previously described in organic solar cells (Figure 3c in ref[55]) and originates from the effect of low mobilities on the apparent shunt and series resistances. For reasonable mobilities and good to very good efficiencies of charge extraction, any reduction in mobility will lead to a slightly reduced shunt resistance (increased shunt conductance), which leads to losses in the *FF*. Once the extraction becomes so poor that the fill factor approaches 25%, any additional reduction in mobility will lead to an additional series resistance that will lead to an increase in the apparent photoshunt (the points lower than $10^{-4}$ cm$^2$V$^{-1}$s$^{-1}$ in Figure 6). However, this situation is mostly of academic relevance, as it only matters for pathologically bad solar cells. Thus, for all situations with good levels of charge extraction, the transport layer mobility directly affects the photoshunt, thereby reducing the $J_{\text{mpp}}$ (photocurrent density at the maximum power point) relative to $J_{\text{sc}}$, leading to a reduction in the fill factor.

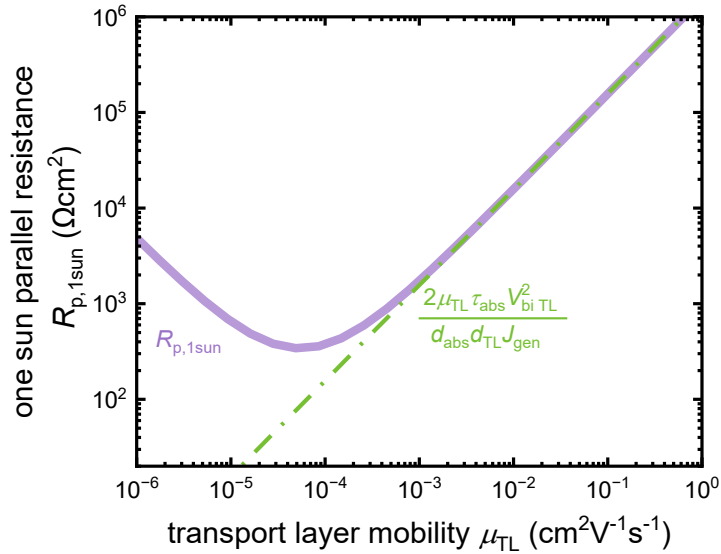

**Figure 6.** Calculated parallel resistance ($R_{p,photo}$) (defined as $R_{p,photo} = dV/dJ|_{V=0}$) under illumination at one sun as a function of the mobility of charge transport layer using equation (13). Higher mobilities generally lead to higher shunt resistances. The exception are extremely low mobilities, where the series resistance becomes so high that the whole current-voltage curve becomes purely ohmic. In this case, the increasing series resistance with lower mobilities also drives up the measured parallel resistance. For relatively efficient devices, however, the shunt resistance and the transport layer mobility are proportional to each other. The green dash-dotted line represents $\frac{2\mu_{TL}\tau_{abs}V_{bi,TL}^2}{d_{abs}d_{TL}J_{gen}}$ which corresponds to the solution of equation (13) if the 1 in the bracket is dominant. This situation is the relevant one for at least moderately efficient solar cells.

It is already clearly revealed that $R_{p,photo}$ increases with the charge transfer layer mobility increasing in Figure 6. Then we will discuss how the charge transfer layer mobility affects solar cell performance. The shifted *J-V* of perovskite-Si tandem cell and perovskite single junction cell under 1 sun illumination based on simulation results with different mobilities are plotted in Figure 7a and b. When the transport layer mobility $\mu_{TL}$ increases from $10^{-6}$ cm$^2$V$^{-1}$s$^{-1}$ to 1 cm$^2$V$^{-1}$s$^{-1}$, the photoshunt feature is observed reduced in both single junction and tandem devices. The simulated efficiencies and *FF*s of tandem cells increase with charge carrier mobility (Figure 7c and d). Once the mobility exceeds $10^{-2}$ cm$^2$V$^{-1}$s$^{-1}$, the *FF*s saturate at about 85% and efficiencies saturate at 38.7%.

Within the limit of the equivalent circuit model of a diode under illumination, the reduction in $J_{mpp}$ of a single junction solar cell (!) as a function of the photoshunt (or any shunt for that matter) can be expressed via[56]

$$J_{mpp} \approx \frac{J_{gen}R_p - V_{mpp}}{R_p + R_s} - J_0 \frac{R_p}{R_p + R_s} \exp\left(\frac{q(V_{mpp} + J_{gen}R_s)}{\left(1 + \frac{R_s}{R_p}\right)n_{id}kT}\right) \quad (14)$$

In the limit of $R_s \ll R_{p,photo}$ that most solar cells will be in, equation (14) simplifies to

$$J_{mpp} \approx \frac{J_{gen}R_p - V_{mpp}}{R_p} - J_0 \exp\left(\frac{q(V_{mpp} + J_{gen}R_s)}{n_{id}kT}\right) \quad (15)$$

In the complete absence of resistive losses, that is for $R_s = 0$, and $R_{p,photo} \to \infty$, equation (15) approaches

$$J_{mpp} \approx J_{gen} - J_0 \exp\left(\frac{qV_{mpp}}{n_{id}kT}\right). \quad (16)$$

Thus, in this ideal case, $J_{mpp}$ is the maximum photocurrent density $J_{gen}$ minus the recombination current density ($J_0 \exp\left(\frac{qV_{mpp}}{n_{id}kT}\right)$) at the maximum power point (MPP), which provides some intuitive meaning to the last three equations. In the presence of a finite photoshunt resistance, $J_{mpp}$ is further reduced by the term $V_{mpp}/R_{p,photo}$ which thereby leads to corresponding losses in fill factor.

The illuminated J-V curves and dark J-V curves of tandem cell, single junction Si and perovskite are plotted on a semi-logarithmic scale in Figure 8. The illuminated J-V curves were shifted to the first quadrant by adding the short-circuit current density $J_{sc}$. The resulting shifted current densities described in Equation (10) be understood as excess recombination current densities relative to the recombination current density present at short circuit. Thus, we can write $J_{rec} = J_{shift} + J_{rec}(V = 0)$. The shifted current density at the maximum power point ($V_{mpp}$) is given by $J_{shift}(V_{mpp}) = J_{sc} - J_{mpp}$ and corresponds to the excess recombination current density at the MPP (dashed green lines in Figure 8). Here, we note that both $J_{sc}$ and $J_{mpp}$ are typically defined (by convention) as positive quantities, i.e. $J_{mpp} = -J(V_{mpp})$, even though the illuminated current-voltage curve of a solar cell is typically drawn into the fourth quadrant of the coordinate system. This value can now be compared to the dark current density at the same voltage. The dark current density is also a recombination current density once the diode region becomes dominant and the current density exceeds the one flowing over the dark shunt resistance. When performing this comparison between the shift current density and the dark current density in Figure 8, we notice that in case of the silicon single junction solar cell, the difference is relatively small, while it is significantly higher in case of the perovskite single junction solar cell. Furthermore, we notice that in case of the Si solar cell, the dark *JV* curve and the shifted *JV* curve are shifted in parallel to each other between approximately 0.55 V to 0.7 V. This parallel shift along the voltage axis corresponds to a constant voltage offset $\Delta V = J_{sc} R_s$ at a given current density that is a characteristic signature of an ohmic series resistance $R_s$.[57, 58, 59] The voltage dependent $R_s$ of the Si solar cell (Figure S6a) is relatively constant as a function of voltage, which proves that the series resistance in this Si solar cell is nearly perfectly ohmic. This is in stark contrast to the perovskite single junction solar cell shown in Figure 8b. Here, no parallel shift is visible, and the shape of the shifted *JV* curve deviates significantly from that of the dark *JV* curve. The $R_s$ of perovskite solar cell (Figure S6b) show highly voltage dependent property compared to Si solar cell.

It is plausible to assume that the recombination current in the dark and under illumination is essentially a function of the Fermi-level splitting inside the perovskite layer and at its interfaces to the transport layers. The fact that for the perovskite solar cell, the excess recombination current under illumination is significantly higher than the recombination current in the dark even without considering the recombination current at short circuit (which is missing in $J_{shift}$) suggests that the Fermi-level splitting is significantly higher at $V_{mpp}$ under illumination than in the dark, whereas this is not the case in the silicon single junction solar cell. This increase of the Fermi-level splitting must be caused by a region of low conductivity in the perovskite solar cell (including its transport layers). This can be understood by considering the equation for the electron current density, which is given by

$$J_n = n\mu_n \frac{dE_{Fn}}{dx}. \tag{17}$$

To drive a current of magnitude $J_{mpp}$ through the perovskite absorber and ETL towards the contact, a gradient of the electron Fermi-level of $dE_{Fn}/dx = J_{mpp}/(n\mu_n)$ is required. This implies that in situations of low conductivity $\sigma_n = qn\mu_n$ as found e.g. in molecular transport layers, the gradient $dE_{Fn}/dx$ must be high, which in turn leads to a higher Fermi-level splitting and a higher recombination current at a given external voltage. The fact that the dark and shifted *JV* curves have different slopes $d\ln(J)/dV$ at any point proves that the difference is not purely ohmic (as in case of the silicon solar cell within a significant range of voltages) but it is non-ohmic throughout and thereby originating from the semiconducting and only moderately conductive regions of the device. This includes the perovskite and the transport layer but importantly excludes the contact layers (e.g. ITO, Ag) that should show purely ohmic conduction due to their conductivity and/or high doping density.

Figure 8c shows the comparison between shifted and dark *JV* curves for the case of a perovskite silicon tandem solar cell. Here, we notice that the strong effect of non-ohmic conduction seen in the perovskite solar cell (panel (b)) has largely disappeared. The voltage dependent $R_s$ of the tandem solar cell (Figure S6c) is relatively constant as a function of voltage which further proves the ohmic conduction in tandem device. This is achieved in practice primarily by moving the tandem cell slightly towards being bottom limited, which implies that the photoshunted region of the perovskite solar cell is cut off by the lower photocurrent that the silicon solar cell generates. This moves the point of maximum efficiency to slightly lower perovskite band gaps as compared to the ideal SQ situation illustrated in Figure 2. The photoshunt of the perovskite solar cell is still visible in the tandem solar cell but it is reduced in such a way that it only matters significantly below $V_{mpp}$. This finding implies that the photoshunt present in perovskite single junction solar cells affects the *FF* of tandem solar cells only indirectly. Higher efficiencies are avoided by band gap and thickness optimization strategies moving away from a situation, where it would ever matter. However, if alternative device designs and transport layers could be designed to overcome the photoshunt issue in perovskite single junction solar cells, it may also be possible to improve tandem solar cell efficiencies by being able to safely move closer to the current matching condition without sacrificing efficiency vs. large *FF* losses (note that some *FF* losses are unavoidable, see Figure 2, but those can be minimized).

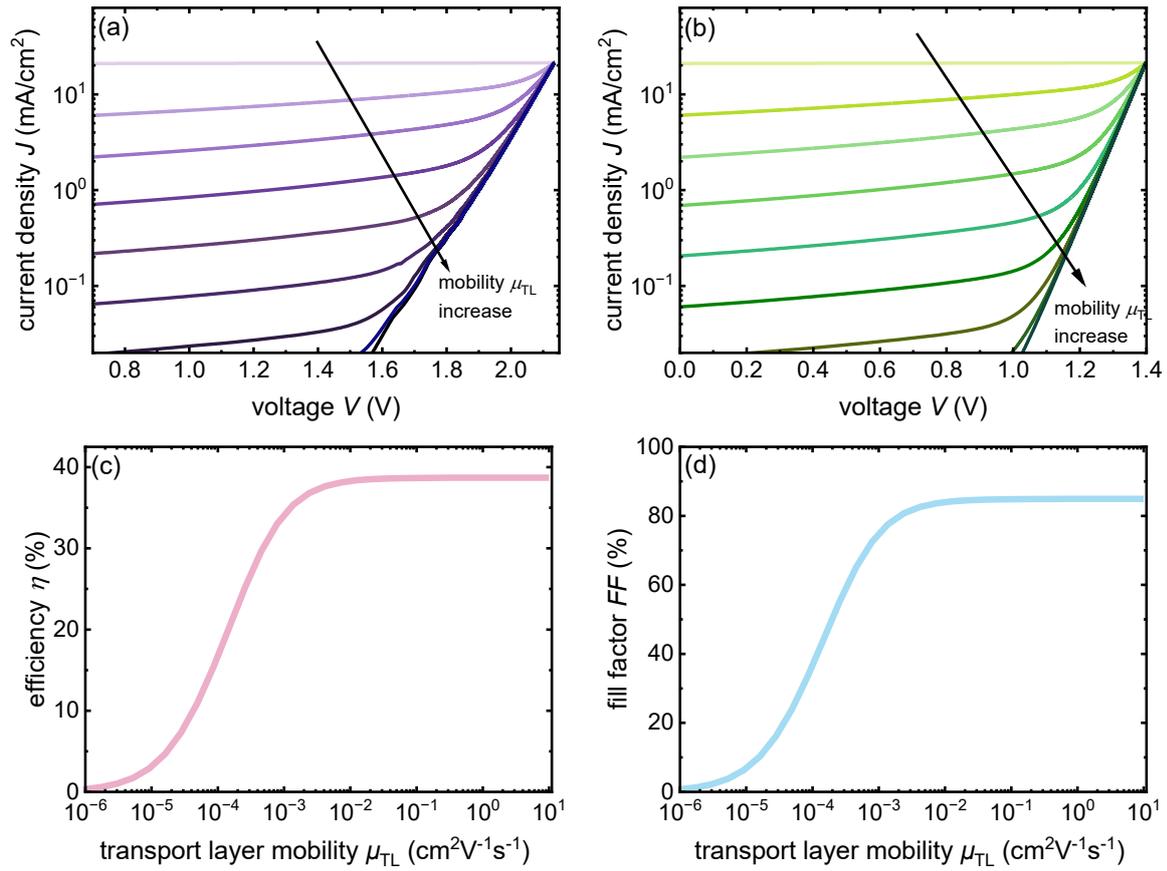

**Figure 7**. Simulated 1 sun illuminated *J-V* of (a) perovskite-Si tandem solar cells and (b) single junction perovskite solar cells shifted to first quandary by adding $J_{sc}$ with varied charge transport layer mobility $\mu_{TL}$ (from $10^{-6}$ to 1 $cm^2V^{-1}s^{-1}$). Simulated (c) efficiencies and (d) *FF*s of perovskite-Si tandem solar cells vs. charge transport layer mobility $\mu_{TL}$ (from $10^{-6}$ to 1 $cm^2V^{-1}s^{-1}$). The bandgap of perovskite is 1.74 eV.

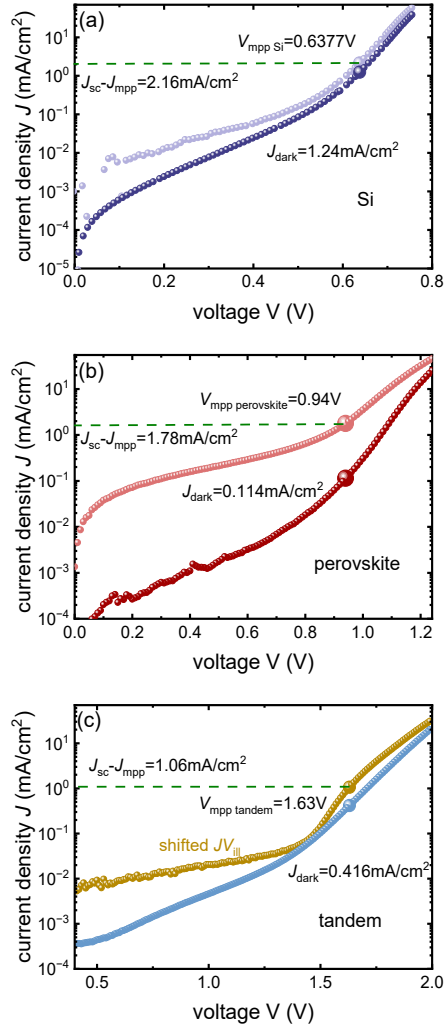

**Figure 8.** Dark current-voltage (*J-V*) curves and illuminated *J-V* curve shifted to the first quadrant by adding $J_{sc}$ of (a) Si single junction solar cell (b) perovskite single junction solar cell and (c) perovskite-Si tandem solar cell.

## 5. Effect of Si bottom cell on the tandem cell fill factor

The shifted illuminated *J-V* curve (yellow in Figure 8c) exhibits a moderate slope in the low-voltage region. This feature is referred to as photoshunt, which indicates poor carrier extraction at low voltages and short circuit. The dark *J-V* curve of the tandem solar cell (blue in Figure 8c) does not exhibit a constant slope similar to that of the single-junction perovskite solar cell (Figure S7a). To further investigate the phenomenon of non-constant slope in dark *J-V* curve in tandem solar cell, we plot the dark *J-V* curves of single junction Si cell (gray) and single junction Si cell (red) in Figure 9a. The dark *J-V* curve of tandem cell in Figure 9a (blue) is obtained by adding up the dark *J-V* of single junction sub-cells. It is obvious that the non-constant slope in dark *J-V* curve comes from Si cell.

The ideality factor $n_{id}$ is valuable for understanding the recombination mechanism and *FF* losses in solar cells. We derived the differential ideality factors of single junction Si and perovskite solar cells as well as tandem solar cell from dark *J-V* curves in Figure 8 according to equation (18) and plotted them in Figure 9b.

$$n_{id} = \frac{q}{kT}\frac{dV}{d\ln(J_d)} \quad (18)$$

*q* is the electron charge. *kT* is the thermal voltage. The differential $n_{id}$ of Si solar cell and perovskite solar cell are highly dependent on voltages and reach the lowest values at voltages around their corresponding $V_{oc}$s. The differential $n_{id}$ of the tandem solar cell taken at its minimum value is approximately equal to the sum of the differential $n_{id}$ of the two sub cells, again taken at their respective minimum value. The edge isolation of Si solar cells was done by laser cutting. There was no additional edge passivation was applied for curing the loss from laser cutting. A PL image of Si bottom cell is shown in Figure S9. We note that the ideality factor of the Si subcell is close to 1 at its respective minimum, but it increases to values above 3 at its peak. This is a somewhat peculiar find as most models of recombination and ideality factors are providing a range between $n_{id}$ = 2/3 for Auger recombination[18] in high-level injection to 2 for SRH recombination via a deep defect. However, there have been previous discussions on ideality factors > 2 in Si solar cells, e.g. by Steingrube et al.[60] Here, the explanation was based on a modified type of SRH recombination that involves interactions between defects, i.e. defect to defect transitions. Whereas this may be a rational explanation for multicrystalline silicon solar cells, it is currently unclear whether similar explanations may explain the high ideality factors in monocrystalline cells. At low voltages, the recombination in the junction dominates and the $n_{id}$ is expected to be around 2. [61, 62] Thus, at the current stage, we merely report the observation of $n_{id}$ > 2 but do not have a plausible interpretation for the finding.

The dark *J-V* of Si solar cell has often been described using a two-diode model which includes two saturation current densities and two ideality factors

$$J_{dark,Si} = J_{01}[\exp(qV/n_{id1}kT) - 1] + J_{02}[\exp(qV/n_{id2}kT) - 1] \ . \quad (19)$$

For the sake of illustrating the effect of the second diode, we assume $n_{id1}$=1 and fix the saturation current density at $J_{01}$=10$^{-12}$ mA/cm$^2$. We vary both the $n_{id2}$ and $J_{02}$ while keeping the value of the recombination current density of the second diode fixed at a voltage of 0.6 V, i.e. $J_{02}[\exp(qV/n_{id2}kT) - 1]$=0.0646 mA/cm$^2$ which corresponds to the value of the dark current density of our measured Si solar cell (in Figure S10). The simulated dark *J-V* of Si solar cells are plotted in Figure 9c. We observe that increasing values of $n_{id}$ lead to increasing current densities at forward voltages < 0.6 V which resembles the effect of lower photoshunt resistances that we discussed previously for the case of perovskite subcells. To study how the different $n_{id}$ of Si bottom cells influence the performance of tandem devices, we simulate the *J-V* curves of perovskite-Si tandem cells under 1 sun illumination and plot them in Figure 9d. The ideal perovskite top cells are simulated in one-diode model with a bandgap of 1.74 eV. We also discussed the case that the perovskite top cell is not ideal in Figure S11. We set the shunt resistance $R_{p,photo}$=10$^6$ Ωcm$^2$ and series resistance $R_s$=0 Ωcm$^2$ to avoid any resistance effects on tandem devices. The $n_{id}$ of Si bottom cells have a non-negligible influence on the performance of perovskite-Si tandem cells, especially $V_{oc}$s and *FF*s as shown in *J-V* curves in Figure 9d. The higher $n_{id}$ values of Si bottom cells cause a reduction in the *FF*s of tandem devices (shown in the insert figure in Figure 9d). However, in the case of the perovskite top cell is a real cell, the *FF* of tandem devices show no difference with Si bottom cells simulated by different $n_{id2}$ (shown in Figure S11).

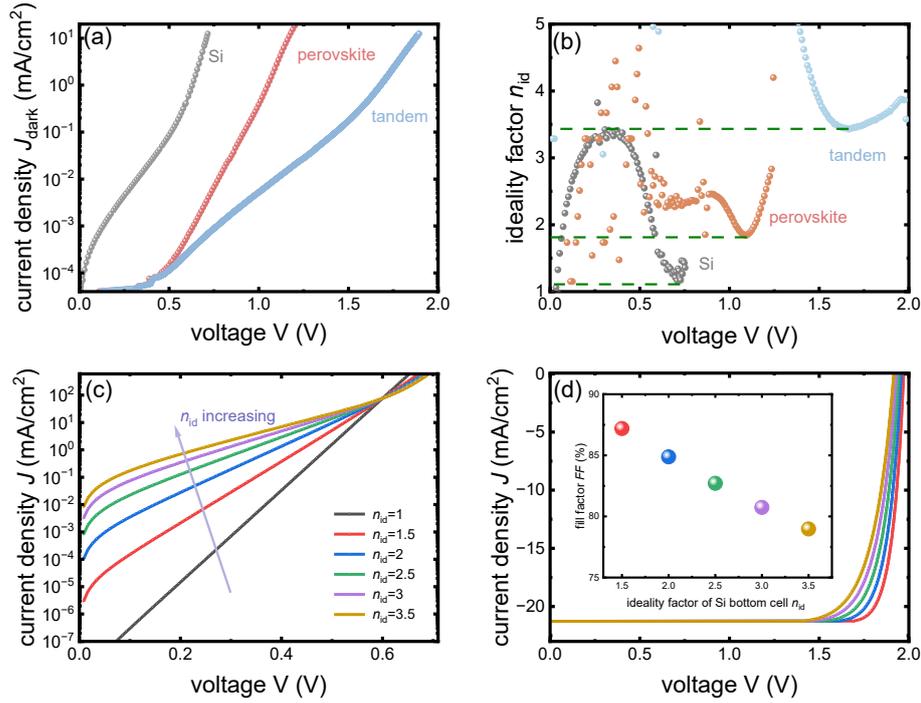

**Figure 9.** (a) Dark *J-V* curve of perovskite-Si tandem cell obtained by adding up measured dark *J-V* curves of single junction Si cell and single junction perovskite cell. (b) differential ideality factors of perovskite-Si tandem solar cell and subcells. (c) Simulated dark *JV* curves of Si solar cells with different ideality factors $n_{id}$. (d) Simulated illuminated *JV* curves of perovskite-Si tandem cells with corresponding Si bottom cells in (c) and ideal perovskite top cells, the fill factors of tandem devices are in the insert figure.

## 6. Effect of Hysteresis and Bias Light on Charge Collection

To study how the current mismatch of individual sub-cells influences the tandem cell, we measured the perovskite-Si tandem solar cell under a sun simulator containing a white LED and an infrared LED. The intensity of the 2 LEDs can be adjusted individually thus the tandem cell can be measured at different current not matching conditions. We kept the short circuit currents of tandem cells at a constant value while changing the intensities of white LED or IR LED. With the intensity of IR light irradiance increasing, the Si sub-cell generates an excess photocurrent not matched by the perovskite solar cell. Thus, the photocurrent of the perovskite solar cell limits the overall photocurrent, thereby leading to an increased visibility of hysteresis features and the photoshunts in the tandem solar cell *J-V* curves. Both features originate from the perovskite subcell and start to influence the tandem solar cell once it is top-cell limited. Figures 10b-d summarize the $V_{oc}$, *FF* and $\eta$ of the tandem solar cell measured at different conditions, where the photocurrents of bottom and top cell are intentionally mismatched. With the white LED irradiation increasing, the perovskite sub-cell is over illuminated, and the tandem cell is in bottom cell limited condition. The $V_{oc}$s of tandem cell are maintained around 1.92V and *FF*s go saturation to about 85%. The hysteresis of the tandem cell is small in this condition. Once the IR LED irradiation increases, the tandem solar cell is in perovskite top cell limited condition. The *FF* and *Voc* of the tandem cell decreased. Assuming that the perovskite subcell within a tandem solar cell typically has similar photoshunts as the devices studied here, it is reasonable to design the tandem cell such that the photocurrent is bottom-cell limited. In this case, the photoshunt would be less visible or invisible in the tandem solar cell even if it did exist in the individual perovskite subcell. If one wants to verify the presence or absence of such a photoshunt, it is possible to make it visible by changing the spectrum in such a way that the Si solar cell produces more photocurrent and the tandem cell becomes top-cell limited.

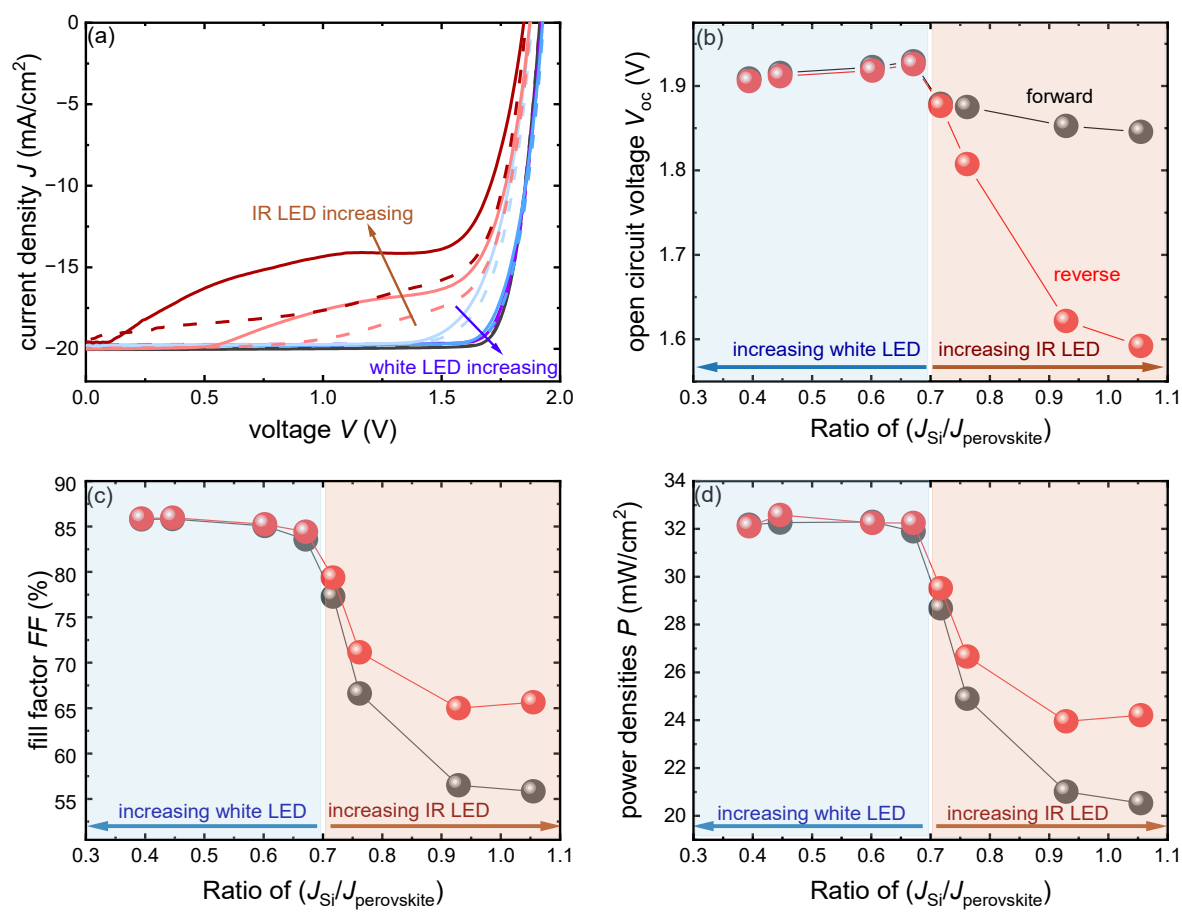

**Figure 10.** (a) Current-voltage (*J-V*) characteristics at different bias light for perovskite-Si tandem cell. The solid lines are forward scan and dash lines are reverse scan. The LED light source contains white and infrared light, and the ratio can be adjusted. Here, we maintain light intensities that make the $J_{SC}$ of the tandem cells equal to the $J_{SC}$ measured at AM 1.5G. The ratio of the two LED light sources varied. The ratio of the two LED lights are described in Table S1, SI (b) $V_{OC}$s (c) *FF*s and (d) output power densities of perovskite-Si tandem cell are plotted

## 7. Conclusions

In summary, this article introduces advanced methods for understanding the fill factor loss mechanisms of monolithic perovskite-Si tandem solar cell by simulation. By comparing the efficiency losses relative to the Shockley-Queisser model of different solar cell technologies, we observe that in perovskite-Si tandem devices, the *FF* is the dominant efficiency source of efficiency losses, similar to other perovskite technology but in contrast to Si single-junctions.[35] The simulation results predict that in the radiative limit, the *FF* of tandem devices has a minimum for the situation of perfect short-circuit current matching. For this reason, highly efficient tandem solar cells are usually designed slightly bottom-cell limited, i.e. a small fraction of the $J_{sc}$ of the tandem cell is sacrificed for a better *FF*. A common strategy to characterize the *FF* loss in monolithic tandem solar cells is extracting pseudo *J-V* curves based on electroluminescence.[30, 63] However, the EL characterization can only isolate the *FF* loss caused by series resistances but cannot explain significant contributions to the *FF* loss in tandem devices that phenomenologically resemble a shunt resistance. In nearly all solar cell technologies, the differential resistance at short circuit is a strong function of light intensity, implying that the shunt visible in a dark JV curve rarely dominates the shape of the JV curve under illumination. These light-intensity dependent apparent shunt resistances (termed here photoshunt) originate from inefficient carrier extraction at short circuit or low forward bias conditions because of the limited mobility of charge transport materials. Thus, the photoshunt is a recombination current created by inefficient extraction that is linear in external voltage but still exponentially dependent on the Fermi-level splitting.[41] Simulation results further confirm that the photoshunts depend primarily on the charge carrier mobilities in the charge transport materials and on the thicknesses of ETL and HTL. These photoshunts in the perovskite subcell can also affect the fill factor of the tandem cell and will do so to different degrees that depend on the exact current matching conditions. In the silicon subcell, we do not find any photoshunts, but a strong second diode with an ideality factor significantly above 2. To investigate whether the observed second diode in Si subcell can affect the *FF* of tandem solar cells, we studied the Si solar cell by two-diode model calculations that were connected in series with the current-voltage curve of the perovskite subcell. Based on the two-diode model simulation, our study further identifies that the non-constant $n_{id}$ of the Si solar cell can slightly affect the *FF* and $V_{oc}$ of tandem devices when the perovskite top cell is assumed to be ideal meaning having an exponential current-voltage curve with ideality factor one and without resistive losses. Finally, we measured the tandem device under different bias light intensities and indicated that the *FF* and efficiency are strongly influenced by spectral changes. The perovskite sub cells usually show a strong photoshunt and suffer from excess charge-carrier recombination at short circuit and forward bias because of the limited mobility of charge transport layers. If the tandem devices are designed slightly bottom-limited or less illuminated in IR region, the effect of photoshunt from perovskite becomes weak because the lower photocurrent generated by Si sub cell cuts off the photoshunt region of the perovskite cell *J-V*. This study broadens the *FF* loss analysis in perovskite-Si tandem solar cell from series resistance relevant issues to charge transport mobility dependent photoshunt. This work also clarifies the synergistic effects of current matching, mobility of transport materials and ideality factors of Si cells on the *FF* of perovskite-Si tandem devices, which points out possible strategies for further device optimization.

**Materials and Methods**

**Materials**

Cesium iodide (CsI), Lead bromide ($PbBr_2$, 98%), N, N'-dimethyl formamide (DMF), dimethyl sulfoxide (DMSO), 2-propanol (IPA), and ethyl acetate were purchased from Thermo Fisher Scientific. Self-assembled monolayers Me-4PACz and MeO-2PACz, bathocuproine (BCP, 99.0%), lead iodide ($PbI_2$, 99.99%), chloroformamidine hydrochloride (Cl-FACl, 98%) and lead chloride ($PbCl_2$, 99.99%) were purchased from TCI. Formamidinium iodide (FAI), Methylammonium chloride (MACl), 4-Fluoro-Phenethylammonium chloride (4F-PEACl) and ethylenediamine diiodide (EDAI) were purchased from Greatcell Solar Materials. F4-TCNQ and $C_{60}$ were from Ossila. Magnesium Fluoride ($MgF_2$, 99.99%) was from Alfa-Aesar. The patterned indium doped tin oxide (ITO) substrates (20mm*20mm) were from KINTEC. All chemicals were used as received.

**Silicon heterojunction solar cell fabrication**

Silicon heterojunction solar cells were fabricated on the double side polished, n-type, (100)-oriented, 4-inch float zone wafers with a thickness of 280 μm and a resistivity of 2 – 5 ohm cm. The texturing process was done in an alkaline-based solution to obtain randomly nano-pyramids. After texturing process, a standard ozone cleaning followed by a native oxide removal process in 1% HF was performed for passivation layer depositions. The hydrogenated intrinsic amorphous silicon a-Si:H (i) and n-type a-Si:H (n) stack was deposited on the front side by PECVD system, with the thickness of approximately 6 nm. Subsequently, a-Si:H (i) and p-type a-Si:H (p) were deposited on the rear side with the thickness of 6 nm and 13 nm, respectively. 70 nm indium tin oxide (ITO) and 1 μm silver were sputtered on rear side through a mask with the aperture area of 1.1 cm × 1.1 cm. On the front side, 20 nm ITO was deposited as the recombination layer with the aligned mask. To recover sputter damage, a heat-assisted light soaking was carried out for 180 s. Finally, the bottom cells were laser scribed into size of 2 cm × 2 cm for tandem fabrication.

**Single junction perovskite solar cell fabrication**

The patterned glass/ITO substrates were cleaned sequentially in Hellmanex III, deionized water, acetone, and 2-propanol for 10 min. Afterwards, the substrates were dried by $N_2$ flow and treated by oxygen plasma at 50 W (Diener Zepto) for 10 min and then transferred to a $N_2$-filled glovebox. The precursor solution of hole transport materials was made by dissolving MeO-2PACz (1 mg/mL) in ethanol and was filtered by a 0.22 μm PTFE filter before use. The solution of MeO-2PACz was deposited on the ITO substrate at 3000 r.p.m for 30 s and then annealed at 100 °C for 10 min.

The 1.69 eV perovskite composition is $Cs_{0.22}FA_{0.78}PbI_{0.85}Br_{0.15}$ with 5% $MAPbCl_3$ and 0.7% Cl-$FAPbCl_3$. $MAPbCl_3$ (the concentration of the solution is 1M) was alloyed by mixing MACl and $PbCl_2$ in DMF: DMSO (1:1) solvent. Cl-$FAPbCl_3$ (the concentration of the solution is 1M) was alloyed by mixing Cl-FACl and $PbCl_2$ in DMF: DMSO (1:1) solvent. The 1.5M perovskite precursor solution for single junction solar cells was prepared by dissolving FAI, CsI, $PbI_2$, and $PbBr_2$ in a solvent mixture of DMF and DMSO with a volume ratio of 5:1 at room temperature, including 85.7 mg CsI, 201.2 mg FAI, 126.3 mg $PbBr_2$ and 546.6 mg $PbI_2$ in 1mL of solvent. The Cl-$FAPbCl_3$ and $MAPbCl_3$ were added to the perovskite precursor individually. The perovskite precursor solution was filtered by a 0.22 μm PTFE filter before use.

The perovskite precursor solution was dropped on substrates and spin-coated at 1000 r.p.m for 10 s then 5000 r.p.m for 40 s. 250 μL ethyl acetate was slightly dropped on the center of the substrate at 38 s after the spin coating start. The perovskite film was annealed on a hot plate at 100 °C for 20 min. To make sure the single junction perovskite solar cells keep stable during characterization, we didn't apply any passivation strategy on the perovskite films. Finally, 20 nm $C_{60}$, 8 nm BCP and 80 nm Ag was deposited sequentially in a K.J.Lesker Mini Spectros System attached to the glovebox (<5×10$^{-6}$ Pa) using a metal shadow mask. The solar cell area is 0.16 cm$^2$.

**Perovskite-Si tandem solar cell fabrication**

The bottom Si solar cells were treated by oxygen plasma at 50 W (Diener Zepto) for 5 min and then transferred to a $N_2$-filled glovebox. The precursor solution of hole transport materials was made by dissolving MeO-2PACz and Me-4PACz (1:1 in weight, 1mg/mL in total) in ethanol. 5 wt% F4-TCNQ was added to the solution of SAMs mixture. The hole transport materials precursor was filtered by a 0.22 μm PTFE filter before use.

The 1.75M perovskite precursor for tandem solar cells was prepared by dissolving a mixture of FAI, CsI, $PbI_2$, and $PbBr_2$ in a solvent mixture of DMF and DMSO with a volume ratio of 5:1, including

100 mg CsI, 234.7 mg FAI, 147.4 mg $PbBr_2$ and 637.7 mg $PbI_2$ in 1mL of solvent. 5% $MAPbCl_3$ and 0.7% $Cl-FAPbCl_3$ were added in the perovskite precursor solution. The perovskite precursor solution was filtered by a 0.45 μm PTFE filter before use.

The perovskite precursor was dropped on the bottom cells and spin coated at 1000 r.p.m for 10 s then 5000 r.p.m for 40 s. 250 μL ethyl acetate was slightly dropped on the center of the substrate at 38 s after the spin coating start. The perovskite film was annealed on a hot plate at 100 °C for 20 min. After the perovskite films cooling down, ethylenediamine diiodide (EDAI) 0.5 mg/mL in IPA was deposited on top of the perovskite film via a dynamic spin coating process of 5000 r.p.m for 30 seconds. 1 mg/mL 4-Fluoro-Phenethylammonium chloride (4F-PEACl) in IPA was also deposited via a dynamic spin coating process of 5000 r.p.m for 30 seconds subsequently. After that, 10 nm $C_{60}$ were deposited by thermal evaporation in a K.J.Lesker Mini Spectros System. $SnO_2$ was deposited by atomic layer deposition (ALD) in a TFS200, Beneq system. The stainless-steel chamber temperature was kept at 80 °C. Tetrakis (dimethylamino) tin (IV) (TDMASn) precursor source was used as Sn precursor and be kept at 55 °C. $H_2O$ source was used as oxidant and be kept at 25 °C. The pulse and purge time for TDMASn was 1 s and 30 s with a carrier gas of high purity nitrogen. The following pulse and purge time for $H_2O$ was 1s and 30s with high purity nitrogen as carrier gas. The two pulse and purge processes were performed for 100 cycles to get about 15nm $SnO_2$. 40 nm IZO as transparent conductive front contact was sputtered from a 3-inch IZO ceramic target on top of the ALD $SnO_2$ layer. The sputtering process was done at 50 W in a mixture of sputter gas (Ar/$O_2$ with 1% vol $O_2$). Ag finger with a thickness of 400 nm was thermally evaporated using a shadow mask. Finally, 100 nm $MgF_2$ was thermally evaporated on top of the IZO layer as an anti-reflection layer.

**Device characterization**

Current-voltage characterization

The illuminated current-voltage (*J-V*) curves of single junction solar cells and tandem solar cells were measured by using a Wavelabs Sinus-70 LED class AAA solar simulator. The sun simulator was calibrated to AM1.5G with an irradiance of 100 mW/cm$^2$ by a calibrated Si reference solar cell of Newport 91150V. The tested solar cells were installed in an electrically contacted measurement box during the *J-V* measurements. A Source Meter from Keithley Instruments (Series 2450) was connected to the measurement box for the *J-V* measurements. The step size of *J-V* measurements is 10 mV.

The dark *J-V* measurements were done with the same process as illuminated *J-V* measurements but without light sources.

The illuminated *J-V* measurements at different bias light intensities were measured under a light source equipped with a white light LED and an infrared LED. The intensities of the two LED light sources can be adjusted individually. The illumination intensity of white LED light source was adjusted to equally 1 sun by matching the integrated short circuit current of the solar cells obtained from EQE measurement. The solar cells were installed in an electrically contacted measurement box and connected to the Keithley source meter (Series 2450). The step size of *J-V* measurements is 10 mV.

External quantum efficiency (EQE)

The EQE measurements of perovskite-Si tandem solar cells were conducted by a setup with a xenon light source (Osram XPO 150 W) and a Bentham monochromator (TMC 300). The spectral range of the EQE setup is from 300 nm to 1100 nm. The simulated light source system was calibrated by using a photodiode (Gigahertz Optik SSO-PD 100-04). The tandem solar cells were installed in an electrically contacted measurement box for EQE characterization. The chopper frequency was 72 Hz. The wavelength step size was 10 nm for the whole measurement. To measure the EQE of perovskite top cell, the Si bottom cell was kept saturated. A bias voltage of 0.7 V was applied to maintain the tandem solar cell at short circuit condition. The EQE of Si bottom cell was measured in a condition that perovskite top cell was saturated by blue light from a LED light source. A bias voltage of 1.2 V was applied on the tandem solar cell to maintain the short circuit condition.

Electroluminescence characterization

The electroluminescence (EL) spectra of perovskite-Si tandem solar cell were recorded by using a home-built electro/photo-luminescence setup. The injected currents were applied on the tandem device with a DC source via an external Keithly SMU 2400 current voltage source which is connected to the EL/PL setup. The tandem solar cell be measured was placed in a measurement box and electrically connected to the Keithly SMU. The injected currents range from 0.3mA to 21mA.

The EL spectra of the tandem solar cell were recorded by a spectrometer (Andor Shamrock 303) equipped with an Andor silicon CCD camera (iDus Series) and an InGaAs detector array (iDus Series). The EL spectra of perovskite top cell at 1.69 eV were recorded by the Andor silicon CCD camera (iDus Series) using a sharp 840 nm short pass filter, while the EL spectra of Si bottom cell at 1.1 eV were recorded by the InGaAs detector array (iDus Series). The EL setup was spectrally calibrated but not calibrated to absolute intensities.

# Supplementary Materials for

# Characterizing Fill Factor Limitations in Perovskite-Silicon Tandem Solar Cells

Yueming Wang[*,1], Nan Sun[1], Chris Dreessen[1], Gaosheng Huang[1], Alexander Eberst[1], Kaining Ding[1], and Thomas Kirchartz[*,1,2]

[1]IMD-3 Photovoltaics, Forschungszentrum Jülich, 52425 Jülich, Germany
[2]Faculty of Engineering and CENIDE, University of Duisburg-Essen, Carl-Benz-Str. 199, 47057 Duisburg, Germany
E-mail: yue.wang@fz-juelich.de
E-mail: t.kirchartz@fz-juelich.de

## Figures and Notes

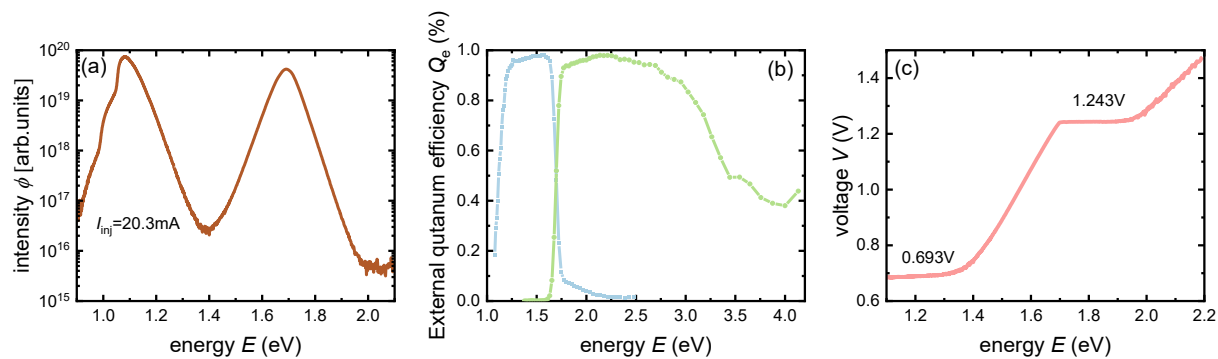

**Figure S1.** (a) electroluminescence spectra of the perovskite and the silicon sub-cell at the injection current 20.3 mA which is equal to short circuit current of tandem cell measured at AM1.5G irradiance. (b) external quantum efficiency spectrum vs. energy of the tandem solar cell. (c) relative internal voltage derived from the EL spectra in Figure S1 (a) according to equation S2.

**Note S1**

Equation (7) can be rewritten as equation (S1) if assuming Boltzmann approximation for the black body photon flux $\phi_{bb}(E)$.

$$\phi_{em}(E) = CQ_e(E)E^2\exp\left(\frac{-E}{kT}\right)\exp\left(\frac{qV}{kT}\right) \tag{S1}$$

$C$ is constant value which needs to be solved later. To solve the internal voltages $V$ of sub-cells, equation (S1) can be rewritten as equation (S2)

$$V = \frac{kT}{q}\ln(\phi_{em}) + \frac{E}{q} - \frac{2kT}{q}\ln(E) - \frac{kT}{q}\ln(Q_e) - \frac{kT}{q}\ln(C) \tag{S2}$$

$Q_e$ is the EQE spectrum of the tandem solar cell shown in Figure S1(b). It needs to be interpolated according to the EL spectrum before applied in equation (S2) to make sure the X-axis are the same. Here the $\phi_{em}$ is the EL spectrum measured at inject current equal to short circuit current 20.3 mA/cm² as shown in Figure S1(a). The EL spectrum of a tandem solar cell can be transferred to internal voltages of the two subcells by using equation (S2). Here the total internal voltages of the two subcells are assumed to be the $V_{oc}$ of the tandem solar cell measured at 1 sun, thus, the constant $C$ in equation (S2) can be solved and used to calculate the internal voltages from EL spectrum at different inject currents. The internal voltages of the two subcells are calculated from the EL spectrum in Figure S1(a) and be plotted in Figure S1(c).

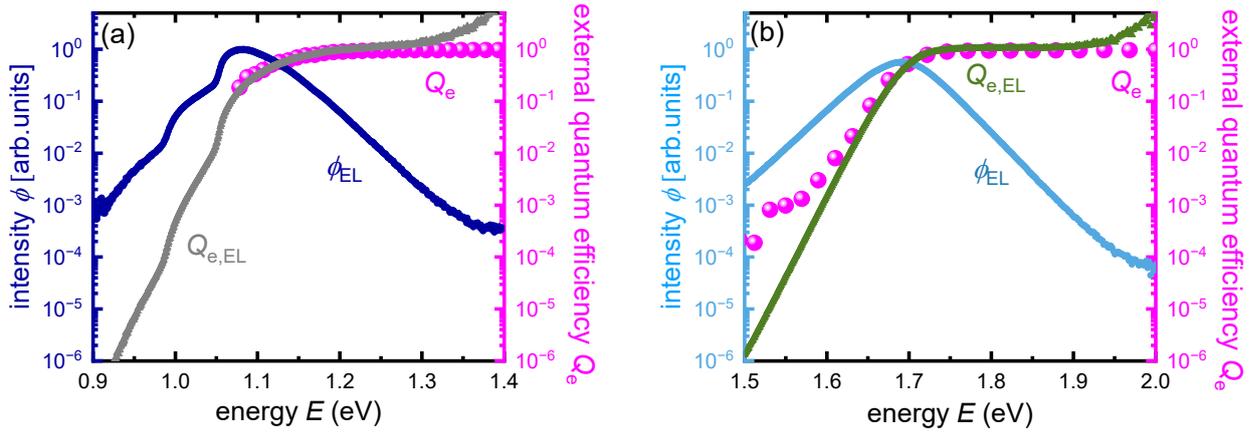

**Figure S2.** The EL spectrum at the injection current equal to short circuit current 20.3 mA/cm² of tandem solar cell, the external quantum efficiency directly measured from solar cell $Q_e$ and external quantum efficiency obtained from EL $Q_{e,EL}$, (a) of the Si bottom cell and (b) of the perovskite top cell.

**Note S2**

To determine the radiative voltages $V_{rad}$ of the two subcells at 1 sun. The saturation value of the radiative recombination current $J_{0,rad}$ for each sub-cell need to be calculated from

$$J_{0,rad} = q \int Q_e \phi_{bb} \, dE \tag{S3}$$

The external quantum efficiency $Q_e$ of the tandem cells is not specific enough as shown in Figure S2 (a) and (b). Accroding to equation (7), a more specific quantum efficiency $Q_{e,EL}$ can be obtained from the EL spectrum measured at injected current $J_{inj}$ =20.3 mA which is equal to the short circuit current of the tandem cell measured at 1 sun irradiance. The $Q_{e,EL}$ can go down to 10⁻⁶ as shown in Figure S2.

The radiative voltage $V_{rad}$ at a given inject current is allowed to be defined by

$$V_{rad} = \frac{kT}{q} \ln(\frac{J_{inj}}{J_{0,rad}}) \tag{S4}$$

The radiative voltage of Si bottom cell $V_{rad,Si}$ is 0.8339V. The radiative voltage of perovskite top cell $V_{rad,perovskite}$ is 1.398V.

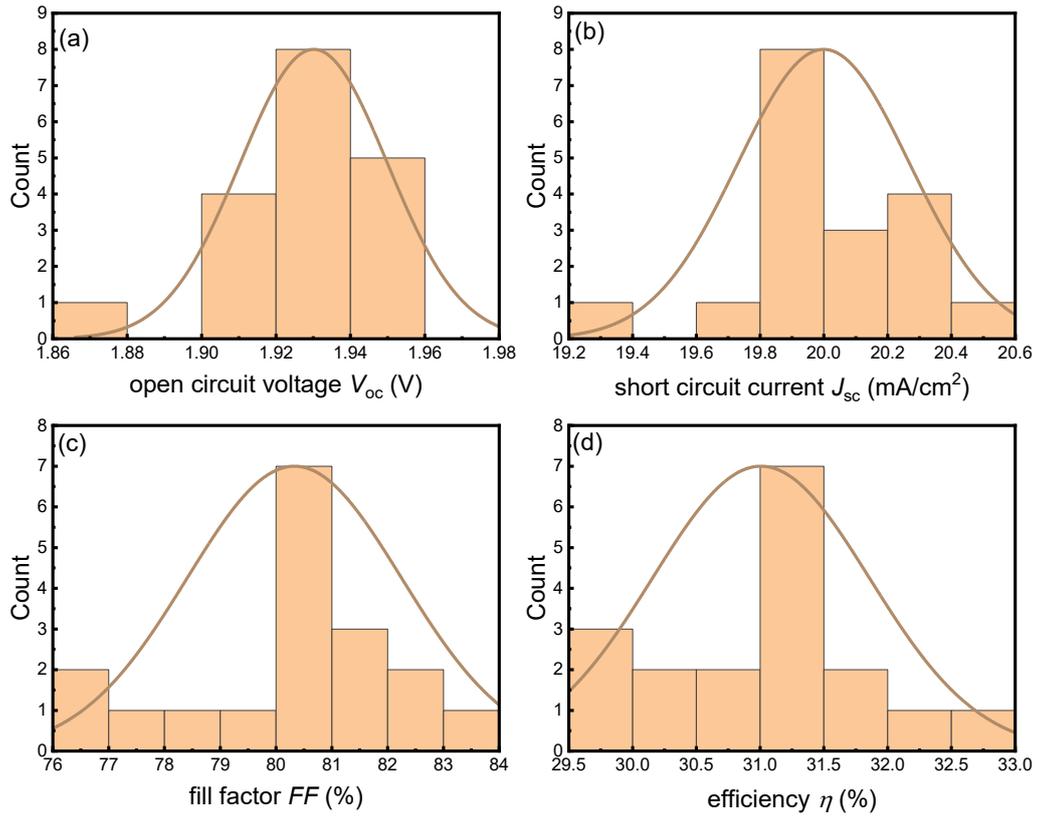

**Figure S3**. The statistics of (a) open-circuit voltage $V_{oc}$, (b) short-circuit current $J_{sc}$ (c) fill factor $FF$, (c) short-circuit current $J_{sc}$ and (d) efficiency for 18 perovskite/Si tandem cells.

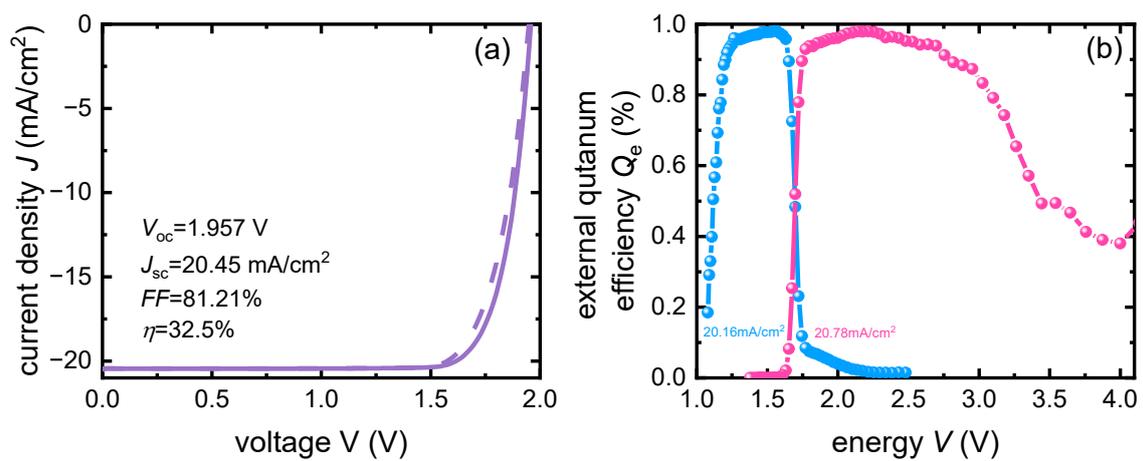

**Figure S4**. (a) current-voltage (*J-V*) curve of the best perovskite-Si tandem solar cell and (b) the external quantum efficiency spectrum.

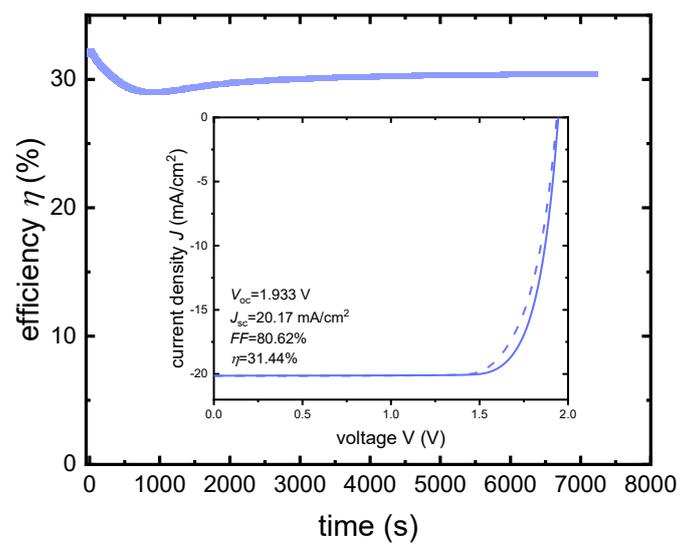

**Figure S5**. Maximum power point tracking of a perovskite/Si tandem solar cell.

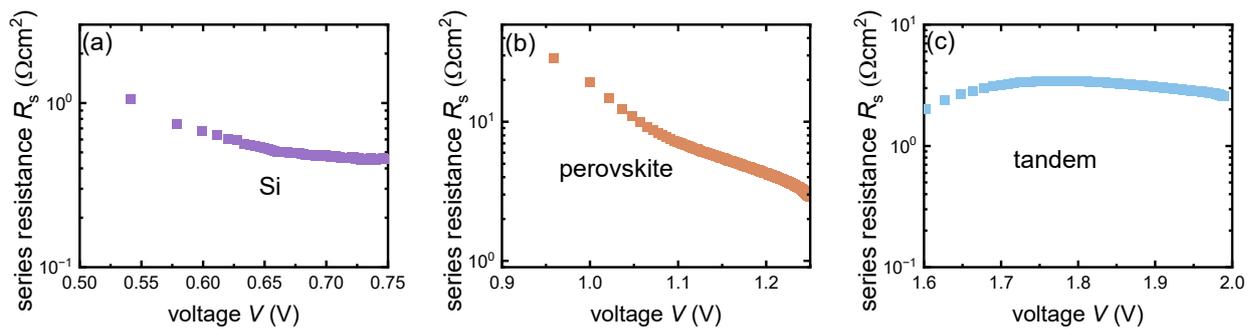

**Figure S6**. Series resistance (a) Si solar cell (b) perovskite solar cell (c) tandem solar cell calculated from the difference between the dark and illuminated *J–V* curve.

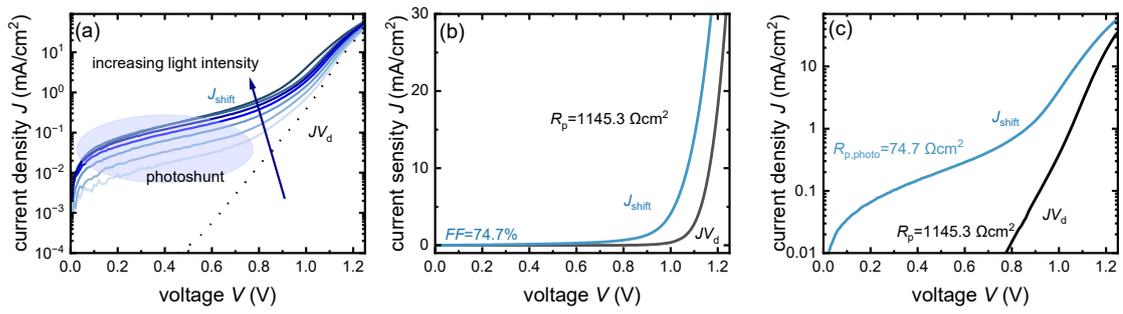

**Figure S7**. (a) semi-log plotted dark current voltage *J-V* curve and illuminated *J-V* curves measured at different light intensities shifted to first quandary according to Equation (10) of a single junction perovskite solar cell. (b) liner plotted (c) semi-log plotted dark current voltage *J-V* curve and illuminated *J-V* curves measured at 1 sun shifted to first quandary according to Equation (10) of a single junction perovskite solar cell.

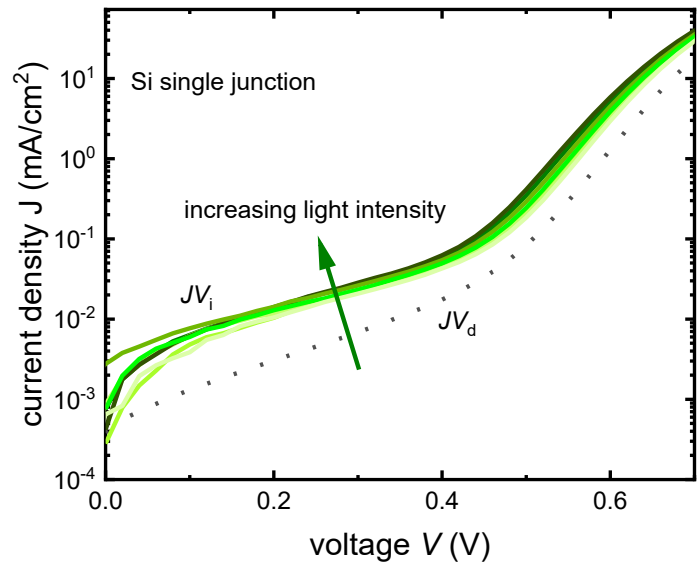

**Figure S8.** Dark current voltage *J-V* curve and illuminated *J-V* curves of a single junction Si solar cell measured at different light intensities.

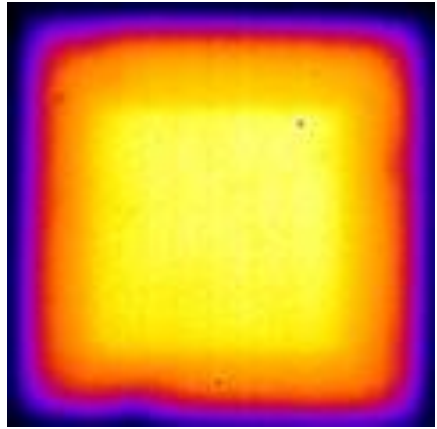

**Figure S9**. Photoluminescence image of a single junction Si solar cell

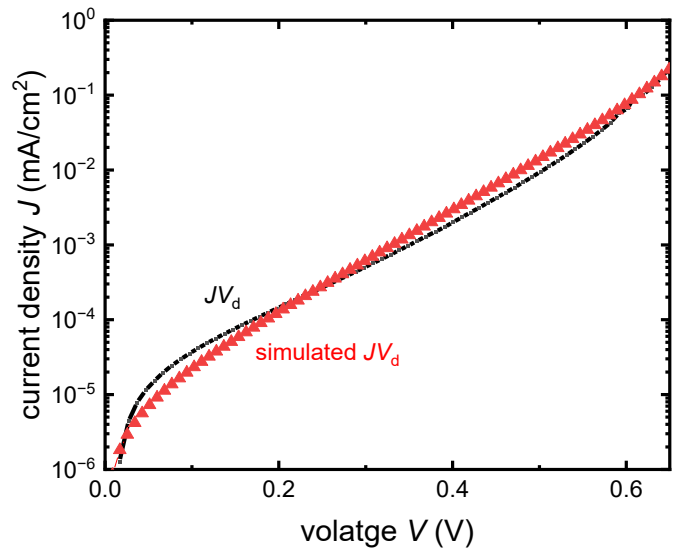

**Figure S10**. Dark current voltage *J-V* curve of a single junction Si solar cell and the simulated dark *J-V* curve according to equation (19).

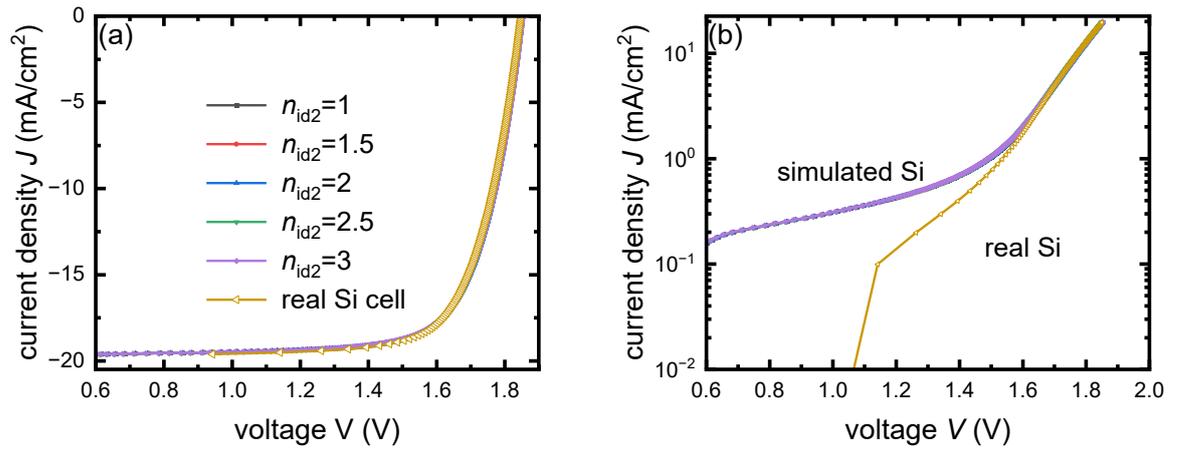

**FigureS11**. Simulated illuminated *JV* curves of perovskite-Si tandem cells combining a real perovskite top cell and different Si bottom cells. The Si bottom cells include the cases of a real Si cell, the Si cells simulated according to equation (19) by two-diode model with different $n_{id2}$. The $n_{id1}$ is assumed to be 1. The $J_{01}$ and $J_{02}$ are obtained by fitting with the dark *JV* of a real Si solar cell with the method in Figure S10. The illuminated *JV* curves of tandem cells are merged from top cell *JV* and bottom cell *JV* by MATLAB and plotted (a) linearly and (b) semi-log.

**Table S1.** The currents driven the LED light sources and the corresponding current density of perovskite subcell or Si sub cell.

| Number | Current of white LED (A) | Current of IR LED (A) | $J_{perovskite}$ (mA/cm$^2$) | $J_{Si}$ (mA/cm$^2$) | Ratio $J_{Si}/J_{perovskite}$ |
|---|---|---|---|---|---|
| 1 | 1.9 | 0.56 | 41.21 | 16.27 | 0.394 |
| 2 | 1.7 | 0.57 | 37.15 | 16.58 | 0.446 |
| 3 | 1.25 | 0.58 | 28.02 | 16.88 | 0.602 |
| 4 | 1.12 | 0.585 | 25.38 | 17.04 | 0.671 |
| 5 | 1.05 | 0.59 | 23.96 | 17.19 | 0.717 |
| 6 | 1 | 0.6 | 22.95 | 17.5 | 0.762 |
| 7 | 0.83 | 0.62 | 19.5 | 18.11 | 0.9287 |
| 8 | 0.83 | 0.7 | 19.5 | 20.57 | 1.055 |